\documentclass[12pt]{iopart}
\usepackage[hidelinks,colorlinks=true,linkcolor=blue, citecolor=blue, urlcolor=blue]{hyperref}
\hypersetup{
  pdftitle={RCM},
  pdfauthor={Tokuro Fukui}
}
\usepackage{cite}
\usepackage{url}
\usepackage{fancybox}
\usepackage{geometry}
\usepackage[dvipdfmx]{graphicx}
\usepackage{caption}
\usepackage{wrapfig}
\expandafter\let\csname equation*\endcsname\relax
\expandafter\let\csname endequation*\endcsname\relax
\usepackage{amsmath,cases}
\usepackage{amssymb}
\usepackage{bbold}
\newcommand{\vect}{\boldsymbol}
\usepackage{tocloft}
\usepackage{mathrsfs}
\usepackage{braket}
\usepackage{longtable}
\usepackage{dcolumn}
\usepackage{multirow}
\usepackage{framed}
\usepackage{arydshln}
\usepackage{chngcntr}
\DeclareSymbolFont{largesymbol}{OMX}{yhex}{m}{n}
\DeclareMathAccent{\Widehat}{\mathord}{largesymbol}{"62}

\usepackage{enumerate}

\usepackage{tikz}
\usetikzlibrary{shapes.geometric}
\usetikzlibrary {shapes.misc}
\usetikzlibrary{positioning}
\usetikzlibrary{arrows.meta}

\usepackage{xcolor}
\usepackage[normalem]{ulem}

\begin{document}

\title[Towards modeling cluster structure of $^8$Be with chiral interaction]{Towards modeling cluster structure of $^8$Be with chiral interaction}

\author{Tokuro Fukui\footnote{Present address: RIKEN Nishina Center, Wako 351-0198, Japan}}

\address{Yukawa Institute for Theoretical Physics, Kyoto University,
Kitashirakawa Oiwake-Cho, Kyoto 606-8502, Japan}
\ead{tokuro.fukui@yukawa.kyoto-u.ac.jp}
\vspace{10pt}
\begin{indented}
\item[]April 2021
\end{indented}

\begin{abstract}
 How the nuclear force behaves in cluster states, in particular those consisting of the $\alpha$ clusters, 
 has been investigated so far, but not yet elucidated.
 Today the chiral effective field theory is established and 
 it would shed new light on the microscopic understanding of the cluster states.
 We aim to address a possible source of the attraction in the cluster states of $^8\mathrm{Be}$
 in view of the pion exchange.
 Namely, we investigate whether the two-pion-exchange interaction acts as a dominant attraction
 in the $\alpha+\alpha$ system as predicted by a previous work.
 We describe theoretically the cluster structure of $^8\mathrm{Be}$ by the Brink model,
 for which the effective interaction is designed from the realistic nuclear force 
 derived through the chiral effective field theory.
 The two-body matrix elements of the chiral interaction with the local-Gaussian bases
 are formulated within the approximation of the spin-isospin saturation forming an $\alpha$ particle.
 Introducing a global prefactor to the chiral interaction phenomenologically,
 the ground and low-lying excited states of $^8\mathrm{Be}$,
 the scattering phase shift of the $\alpha$-$\alpha$ system as well, are satisfactorily depicted.
 The attraction in the cluster states is found to be stemming from the two-pion-exchange contributions dominantly,
 along with nonnegligible short-range terms.
 The present work can be the foundation towards constructing realistic cluster models,
 by which the cluster states will be revealed microscopically in the next step.
\end{abstract}

%
%
%
%
%

\section{Introduction}
\label{SecIntro}
Clustering is one of the fundamental aspects of many-body nuclear systems.
With the aim of understanding cluster states of light nuclei in view of realistic nuclear force,
an enormous number of theoretical studies have been carried out so far.
For instance, there have been approaches to the microscopic description of $\alpha$-cluster structure and its scattering based on
the Brueckner generator coordinate method (GCM)~\cite{10.1143/PTP.44.646,10.1143/PTP.45.1515,10.1143/PTP.59.817},
the Brueckner antisymmetrized molecular dynamics~\cite{10.1143/PTP.117.189,10.1143/PTP.121.299,10.1143/PTP.124.315,Kato12},
the quantum Monte Carlo methods~\cite{PhysRevC.62.014001,RevModPhys.87.1067,PhysRevLett.111.062502},
the no-core shell model (NCSM)~\cite{PhysRevLett.84.5728,PhysRevC.62.054311,PhysRevC.86.054301,10.1093/ptep/pts012,PhysRevLett.114.212502,PhysRevLett.119.062501},
the symmetry-adapted NCSM~\cite{PhysRevLett.98.162503,PhysRevLett.111.252501,PhysRevC.102.044608},
the NCSM with the resonating-group method~\cite{PhysRevLett.113.032503,kravvaris2020ab},
the NCSM with continuum~\cite{PhysRevLett.117.222501},
the fermionic molecular dynamics~\cite{NEFF2004357,PhysRevLett.98.032501},
and the nuclear lattice effective field theory (EFT)~\cite{Borasoy2007-1,Borasoy2007-2,PhysRevLett.106.192501,PhysRevLett.109.252501,Epelbaum2013,PhysRevLett.112.102501,Elhatisari2015,PhysRevLett.117.132501,PhysRevLett.119.222505}.
We also refer the readers to a recent review paper~\cite{RevModPhys.90.035004} on related topics.

In spite of intensive studies by the above works, 
the mechanism how nucleons gain the attraction in the $\alpha$-cluster states needs to be further clarified.
In Ref.~\cite{10.1143/PTP.27.793} the two-pion ($2\pi$)-exchange interaction was assumed to be a source of the attraction
in the $\alpha$-cluster states of $^8\mathrm{Be}$,
relying on the fact that there is no attraction stemming from the one-pion ($1\pi$) exchange in the direct term (Hartree term)
between nucleons, each of which belongs to different $\alpha$ particles.
However, this assumption has not been confirmed yet.
In the 1960s, when Ref.~\cite{10.1143/PTP.27.793} was published,
the dynamics of nucleons and pions had not been fully revealed.
Today, we understand that such dynamics is constrained by the chiral symmetry,
and the chiral EFT~\cite{Weinberg1979327,Epelbaum2006654,MACHLEIDT20111} based on this symmetry
controls the pion-exchange contributions of the nuclear force in a hierarchical way.
Thus, we expect the above assumption can be inspected by employing the chiral interaction.

An important concept to characterize the $\alpha$ clustering is ``internally strong but externally weak''~\cite{10.1143/PTPS.52.1,10.1143/PTPS.52.25},
indicating that each of $\alpha$ particles is bound strongly but they weakly interact with each other.
The Volkov interaction~\cite{VOLKOV196533}, a phenomenological effective interaction commonly employed,
realizes the above nature, namely ``externally weak'', by introducing the unnaturally strong odd-state repulsion.
In contrast, by a microscopic approach of the Brueckner-GCM~\cite{10.1143/PTP.44.646},
the property ``externally weak'' was explained by the realistic tensor force, 
which becomes weaker in $^8\mathrm{Be}$ compared to that in $^4\mathrm{He}$.
Thus, the saturation mechanism was naturally understood.
It is intriguing to investigate the above concept from the view point of the chiral EFT.

In the present work, we concentrate on the typical cluster states of $^8\mathrm{Be}$ in the $\alpha+\alpha$ configuration,
and aim to address which terms of the chiral interaction 
(the contact, $1\pi$-exchange, and/or $2\pi$-exchange contributions) 
account for the origin of the attraction in the $\alpha$-cluster states, using a simple model.
To this end, we adopt the Brink model~\cite{brink1966proc} presuming the spin-isospin saturation
forming an $\alpha$ particle.
The effective interaction relevant to the Brink model is phenomenologically prepared 
from the realistic nucleon-nucleon interaction of the chiral EFT
at next-to-next-to-next-to-leading order (N$^3$LO),
where a high-precision nuclear potentials are available~\cite{ENTEM200293,PhysRevC.66.014002,PhysRevC.68.041001,MACHLEIDT20111}.

We emphasize that the relative strengths among 
the contact, $1\pi$-exchange, and $2\pi$-exchange contributions, are expected to be unchanged
in the effective interaction in the present work.
This is the reason why we employ a phenomenological approach,
which is sufficient to fulfill our purpose that 
we address only the source of the attraction in the $\alpha+\alpha$ system.
Of course, what we can elucidate in this paper is limited.
Therefore, we regard this work as the first step towards fully microscopic approaches,
by which the whole picture of the nuclear force in the clustering is expected to be revealed.
Indeed, the two-body matrix elements (MEs) of the chiral interaction at N$^3$LO
within the local-Gaussian states formulated here is applicable to realistic cluster models
such like the Brueckner GCM.

This article is organized as follows.
Section~\ref{TheorFrame} is devoted to the overview of the theoretical framework,
namely the Brink model and the effective Hamiltonian.
In Sec.~\ref{ResDis}, the calculated energies, radii, and phase shifts
are compared with the measured data.
We also argue the possible source of the attraction in the $\alpha$-cluster states.
A summary and future perspectives are given in Sec.~\ref{Summary}.
We show in \ref{SecParaDep} that our conclusion is robust against the variation of parameters we use.
The formalism of the two-body MEs is relegated to \ref{SecMEs}.
The constants we adopt are listed in \ref{SecConstants}.

\section{Theoretical framework}
\label{TheorFrame}
\subsection{Brink model}
\label{TheorFrameBrink}
We describe the cluster states by the Brink model~\cite{brink1966proc},
where the single-particle wave function is given by
\begin{align}
 \braket{\vect{r} | \phi_i \chi_a }
 &=
 \phi_i(\vect{r})
 \ket{\chi_a}
 =
 \left(\frac{2\nu}{\pi}\right)^{\!\!\frac{3}{4}}
 \exp\!\!\left[-\nu\left(\vect{r}-\vect{R}_i\right)^2\right]
 \ket{\chi_a}.
 \label{spwf1}
\end{align}
Here $\vect{r}$ is the position of nucleon, while $\vect{R}_i$ is the position of the $\alpha$ clusters,
and $\nu$ determines the range of the Gaussian.
The index $a$ specifies the spin-isospin states;
\begin{align}
 \ket{\chi_a}
 =
 \begin{cases}
  \ket{\uparrow_{\sigma}\uparrow_{\tau}}     &\qquad\left(a=1\right),\\
  \ket{\uparrow_{\sigma}\downarrow_{\tau}}   &\qquad\left(a=2\right),\\
  \ket{\downarrow_{\sigma}\uparrow_{\tau}}   &\qquad\left(a=3\right),\\
  \ket{\downarrow_{\sigma}\downarrow_{\tau}} &\qquad\left(a=4\right),
 \end{cases}
 \label{spwf2}
\end{align}
where $\uparrow_{\sigma}$ and $\downarrow_{\sigma}$ stand for the spin-up and spin-down states, respectively,
and similar to the isospin with the symbol $\tau$.
Within the Brink model, the $\alpha$ cluster is presumed as the spin-isospin-saturated four nucleons,
i.e., the $(0s)^4$ configuration is adopted.
Naturally, the Brink model is tailored for systems described as $\alpha$ clusters only.
In such an ansatz, the many-body wave function is given by the Slater determinant as
\begin{alignat}{2}
 \ket{\Psi_{\mathrm{Brink}}}
 &&=\sqrt{A!}\hat{\mathcal{A}}_{A}
 &\left[\ket{\phi_1\chi_1} \, \ket{\phi_1\chi_2}  \, \ket{\phi_1\chi_3} \, \ket{\phi_1\chi_4} \right.
 \nonumber\\
 &&&\left. \ket{\phi_2\chi_1} \, \ket{\phi_2\chi_2} \, \ket{\phi_2\chi_3} \, \ket{\phi_2\chi_4} \right.
 \nonumber\\
 &&&\qquad\qquad\vdots
 \nonumber\\
 &&&\left. \ket{\phi_{N_\alpha} \chi_1} \, \ket{\phi_{N_\alpha} \chi_2} \, \ket{\phi_{N_\alpha} \chi_3} \, \ket{\phi_{N_\alpha} \chi_4} \right],
 \label{mbwfBrink}
\end{alignat}
with the number of the $\alpha$ particles $N_{\alpha}=A/4$.
Here, four nucleons are located in a point of the coordinate space by sharing the
same Gaussian center $\vect{R}_i$, while their spin and isospin states
are different with each other to form an $\alpha$ particle.
The $A$-body antisymmetrizer $\hat{\mathcal{A}}_{A}$, which is a projection operator, 
is expressed symbolically as
\begin{align}
 \hat{\mathcal{A}}_{A}
 =
 \frac{1}{A!}\sum_P (-)^P \hat P.
 \label{Aantisym}
\end{align}
The summation over $P$ results in $A!$ terms, and
$(-)^P$ is $1$ or $-1$ depending on the even or odd permutation respectively.
The permutation operator $\hat P$ is symbolic of any possible
permutation.
For example, the two- and three-body antisymmetrizers can be respectively written by
\begin{align}
 \hat{\mathcal A}_2
 &=
 \frac{1}{2}\left(\mathbb{1}-\hat{\mathcal{P}}_{12}\right),
 \label{antisymope2b_again}\\
 \hat{\mathcal A}_3
 &=
 \frac{1}{3!}\left(\mathbb{1}-\hat{\mathcal{P}}_{12}-\hat{\mathcal{P}}_{23}-\hat{\mathcal{P}}_{31}
 +\hat{\mathcal{P}}_{12}\hat{\mathcal{P}}_{23}+\hat{\mathcal{P}}_{12}\hat{\mathcal{P}}_{31}\right),
 \label{antisymope3b}
\end{align}
where $\hat{\mathcal{P}}_{12}$ is the permutation operator exchanging the particles $1$ and $2$.

The many-body states $\ket{\Psi_{\mathrm{Brink}}}$ are projected onto the eigenstates of the angular momentum $J$ as
\begin{align}
 \ket{\Psi_{MM'}^J}
 &=
 \frac{2J+1}{8\pi^2}
 \int d\Omega D_{MM'}^{J*}(\Omega)\hat R(\Omega)
 \ket{\Psi_{\mathrm{Brink}}},
 \label{JMMstate}
\end{align}
where $D_{MM'}^{J}$ is the Wigner $D$-function with the eigenvalues $M$ and $M'$ of the $z$-component of $J$,
while $\hat R$ is the rotation operator acting on the coordinate and spin states.
The integration over the three-dimensional Euler angle $\Omega$ is performed numerically.

As regards $^8\mathrm{Be}$, two $\alpha$ clusters can be settled in a one-dimensional configuration,
namely $\vect{R}_1=d\vect{e}_z/2$ and $\vect{R}_2=-d\vect{e}_z/2$ with the unit vector $\vect{e}_z$
of the three-dimensional Cartesian coordinates.
When the Hamiltonian is given, the corresponding energy eigenvalues can be obtained 
as a function of the $\alpha$-$\alpha$ relative distance $d$
within the limit of the single-Slater determinant.

In order to describe the system more precisely, we superpose the Slater determinants
associated with different values of $d$ based on the GCM~\cite{PhysRev.89.1102,PhysRev.108.311,brink1966proc} as
\begin{align}
 \ket{\Psi_{MM'}^J}_{\mathrm{GCM}}
 =
 \sum_k c_k
 \ket{\Psi_{MM'}^J(d_k)}.
 \label{GCMwf}
\end{align}
Here the subscript $k$ of $d$ is shown explicitly.
The coefficient $c_k$ is obtained by solving the Hill-Wheeler equation~\cite{PhysRev.89.1102,PhysRev.108.311,brink1966proc}.
The phase shift of the $\alpha$-$\alpha$ scattering is also calculated
based on the Kohn-Hulth\'en variational principle with the GCM~\cite{10.1143/PTP.56.583,10.1143/PTPS.62.236}.

Since the $\alpha$-cluster structure is presumed,
we cannot discuss the intrinsic structure of $^4\mathrm{He}$ by the Brink model.
Therefore, the calculated energy of $^4\mathrm{He}$ is always subtracted from that of $^8\mathrm{Be}$,
as explained in Sec.~\ref{ResDis1}, 
and thus we address the relative energy of the $^8\mathrm{Be}$-ground state measured from the $\alpha+\alpha$ threshold.

\subsection{Hamiltonian}
\label{TheorFrameHamil}
The many-body Hamiltonian in the present framework is given by
\begin{align}
 \hat H
 &=
 \sum_{i=1}^A \hat T_i-\hat T_G
 +\sum_{i<j}^A \left[\hat V_{ij}^{(\mathrm{N})}+\hat V_{ij}^{(\mathrm{C})}\right],
 \label{mbHamil}
\end{align}
where $\hat T_i$ is the kinetic-energy operator of the $i$th nucleon
and that of the center-of-mass contribution $\hat T_G$ is subtracted.
We here consider only the two-body interaction
consisting of the nuclear interaction $\hat V_{ij}^{(\mathrm{N})}$ and the Coulomb interaction $\hat V_{ij}^{(\mathrm{C})}$.

We employ the chiral interaction at N$^3$LO~\cite{ENTEM200293,PhysRevC.66.014002,PhysRevC.68.041001,MACHLEIDT20111} 
as $\hat V_{ij}^{(\mathrm{N})}$, which can be classified into seven terms as
\begin{align}
 \hat V_{ij}^{(\mathrm{N})}
 &=
 \hat V_{\mathrm{ct}}^{(0)}+\hat V_{1\pi}^{(0)}+\hat V_{\mathrm{ct}}^{(2)}+\hat V_{2\pi}^{(2)}
 +\hat V_{2\pi}^{(3)}+\hat V_{\mathrm{ct}}^{(4)}+\hat V_{2\pi}^{(4)}.
 \label{chiralV}
\end{align}
Using the chiral-expansion power $n_\chi$, the superscripts stand for the interaction at leading order (LO)
with $n_\chi=0$, next-leading order (NLO) with $n_\chi=2$, next-to-next-to-leading
order (N$^2$LO) with $n_\chi=3$, and N$^3$LO with $n_\chi=4$.
The $2\pi$-exchange interaction $\hat V_{2\pi}^{(n_\chi)}$ and the contact interaction $\hat V_{\mathrm{ct}}^{(n_\chi)}$
depend on $n_\chi$,
while the $1\pi$-exchange interaction $\hat V_{1\pi}$ contributes only at LO.

Then we need to derive the effective Hamiltonian $\hat H_{\mathrm{eff}}$ 
relevant to the Brink model from the realistic Hamiltonian $\hat H$.
In order to be consistent with the concept of constructing realistic nuclear forces, 
a microscopic way to derive $\hat H_{\mathrm{eff}}$ such as the Brueckner-GCM approach is necessary.
However, in this paper, we obtain $\hat H_{\mathrm{eff}}$ in a phenomenological way
as the first step towards the realistic cluster model based on the Brueckner theory.
We introduce the prefactor $c_0$, and thus $\hat H_{\mathrm{eff}}$ is given by
\begin{align}
 \hat H_{\mathrm{eff}}
 &=
 \sum_{i=1}^A \hat T_i-\hat T_G
 +\sum_{i<j}^A \left[\hat V_{ij}^{(\mathrm{N;\,eff})}+\hat V_{ij}^{(\mathrm{C})}\right],
 \label{mbHamileffc0}\\
 \hat V_{ij}^{(\mathrm{N;\,eff})}
 &=
 c_0 \hat V_{ij}^{(\mathrm{N})}.
 \label{Veffc0}
\end{align}
The prefactor $c_0$ is fixed so that the relative energy of the $^8\mathrm{Be}$-ground state measured from the $\alpha+\alpha$ threshold obtained with the GCM coincides with the experimental value.

Here we emphasize that, in spite of the introduction of the phenomenological prefactor globally, 
the relative strength of each term of the chiral interaction remains realistic,
i.e., the low-energy constants (LECs) are unchanged from those of the realistic force.
Therefore we can address which term of the interaction in view of the pion exchange 
plays an essential role on the cluster states.
In \ref{SecParaDepc0} we show that our conclusion is robust against the variation of $c_0$ from unity to a certain value
fixed by the strategy mentioned above.

In the Brink model, due to the ansatz of the spin-isospin saturation with respect to the $\alpha$ particle,
the antisymmetrized two-body MEs of the noncentral terms involved in $\hat V_{ij}^{(\mathrm{N})}$ become zero.
In \ref{SecMEs} we formulate the two-body MEs of the chiral interaction at N$^3$LO within the Brink model.
The formalism of the two-body MEs is applicable to the Brueckner-Brink/GCM model, 
which will be constructed in future.

As mentioned in Sec.~\ref{SecIntro}, the effective interaction derived from the Brueckner theory,
which can handle the tensor force in a straight-forward way, 
is desirable to account for the concept ``internally strong but externally weak''.
In the present approach, however, the introduction of $c_0$ may not be adequate to explain this concept, namely ``internal strong''.
Instead, this paper provides a basis towards such a fully microscopic approach.
Therefore, the derivation of the effective interaction by the Brueckner theory starting from the chiral-EFT interaction
will be a vital work in future, and thus, the tensor-force effect and three-nucleon force of the chiral interaction
on the clustering will be clarified.

Furthermore, we comment on the noncentral contribution to the cluster states of $^8\mathrm{Be}$, in particular the tensor-force effect.
In Refs.~\cite{PhysRevC.98.054306,10.1093/ptep/ptz046}, within the framework of the antisymmetrized quasicluster model,
the authors have shown that the tensor force plays a dominant role inside $^4\mathrm{He}$,
corresponding to the shift of the $2\alpha$'s-threshold energy.
It results in the tensor-force effect acting repulsively in the relative energy of the $^8\mathrm{Be}$-ground state measured from the threshold, 
when the two clusters approach.

\subsection{Model setup}
\label{TheorFrameModel}
The size parameter $\nu$ is fixed so that the calculated root-mean-square (RMS) matter radius $R_\mathrm{M}$ of $^4\mathrm{He}$
is consistent with the observed value $R_\mathrm{M}^{(\mathrm{exp})}~\!\!\!=~\!\!\!1.45$ fm.
Note that $R_\mathrm{M}^{(\mathrm{exp})}$ is extracted from 
$R_\mathrm{M}^{(\mathrm{exp})}~\!\!\!=~\!\!\!\sqrt{R_\mathrm{C}^2-R_p^2-(N/Z)R_n^2}$
with the proton and neutron numbers $Z$ and $N$, respectively, 
the measured charge radius $R_\mathrm{C}=1.68$~fm~\cite{ANGELI201369} of $^4\mathrm{He}$,
and the squared proton (neutron) charge radius, experimentally obtained as
$R_p^2=0.832$~fm$^2$~\cite{BORISYUK201059} ($R_n^2=-0.115$~fm$^2$~\cite{ANGELI201369}).
Choosing $\nu=0.26~\mathrm{fm}^{-2}$, corresponding to the harmonic-oscillator energy, $\hbar\omega = 2\hbar^2 \nu/m_N\sim22$~MeV,
we compute $R_\mathrm{M}$ as 1.47~fm, where $m_N$ is the average nucleon mass.
Although the choice of $\nu=0.26~\mathrm{fm}^{-2}$ gives the proper $R_\mathrm{M}$,
our model cannot result in the reasonable value of the $^4\mathrm{He}$-ground-state energy at the same time, as shown in Sec.~\ref{ResDis}.
Therefore, in \ref{SecParaDepnu}, we investigate the sensitivity to $\nu$ in the description of the $^8\mathrm{Be}$-cluster state,
and confirm the validity of the current value of $\nu$.

We test three sets of the parameterization of the chiral interaction, depending on the cutoff $\Lambda$,
namely, the widely adopted value, $\Lambda=500$~MeV~\cite{PhysRevC.68.041001,MACHLEIDT20111}, 
as well as $\Lambda=450$~\cite{PhysRevC.87.014322,PhysRevC.89.044321} and 600~MeV~\cite{MACHLEIDT20111}.
The physical constants and LECs we use are summarized in \ref{SecConstants}.
The prefactor $c_0$, which is determined from the comparison of the calculated relative energy of the $^8\mathrm{Be}$-ground state
based on the GCM with the measured value, also depends on $\Lambda$.
Thus, all the calculations below are carried out with 
$c_0=1.37,~1.79,$ and $2.15$ for $\Lambda=450, ~500,$ and $600$~MeV, respectively, as listed in Table~\ref{tablec0}.
See also \ref{SecParaDepc0}, where we show how the results vary when $c_0$ is evolved from unity to the value given in Table~\ref{tablec0}.

The GCM calculations are performed with the twelve Slater determinants (the maximum value of $k$ is 12)
specified by $d_k$ from 1.0 to 8.7~fm with the 0.7-fm interval.
The same basis functions are applied to the phase-shift calculations.
\begin{table}[!t]
 \caption{The prefactor $c_0$ fixed by the comparison of the GCM-calculated relative energy of the $^8\mathrm{Be}$-ground state with measured values.}
 \label{tablec0}
 \begin{center}
   \begin{tabular*}{0.7\textwidth}{@{\extracolsep{\fill}}c|ccc}
    \hline
    \multirow{2}{*}{$c_0$}  & $\Lambda=450$~MeV & $\Lambda=500$~MeV & $\Lambda=600$~MeV \\
                            & $1.37$            & $1.79$            & $2.15$            \\
    \hline
   \end{tabular*}
 \end{center}
\end{table}

\section{Results and discussion}
\label{ResDis}
\subsection{Energies, radii, and phase shifts}
\label{ResDis1}
In Table~\ref{tableEalpha}, we list the ground-state energy of $^4\mathrm{He}$, denoted by $E_\alpha$, computed with the present setup.
The calculated values of $E_\alpha$ are not far from the experimental data~\cite{Wang_2021}, except for the case of $\Lambda=600$~MeV.
Note that, $\hat H_{\mathrm{eff}}$ with $c_0$ fixed for the ground state of $^8\mathrm{Be}$ 
does not give the optimal $E_\alpha$ at around $\nu=0.26$~fm$^{-2}$.
Therefore, the description of $^4\mathrm{He}$ must be improved in a forthcoming work,
where $\hat H_{\mathrm{eff}}$ will be derived in a fully microscopic way.
\begin{table}[!t]
 \caption{The ground-state energy of $^4\mathrm{He}$ calculated with $\hat H_{\mathrm{eff}}$ defined by Eq.~\eqref{mbHamileffc0}.
 The experimental value given in the rightmost column is taken from Ref.~\cite{Wang_2021}.}
 \label{tableEalpha}
 \begin{center}
   \begin{tabular*}{0.95\textwidth}{@{\extracolsep{\fill}}c|cccc}
    \hline
    \multirow{2}{*}{$E_\alpha$ (MeV)}  & $\Lambda=450$~MeV & $\Lambda=500$~MeV & $\Lambda=600$~MeV & Exp \\
                                       & $-33.0397$        & $-34.3761$        & $-58.6637$        & $-28.2957$\\
    \hline
   \end{tabular*}
 \end{center}
\end{table}

\begin{figure}[!b]
\begin{center}
\includegraphics[width=0.70\textwidth,clip]{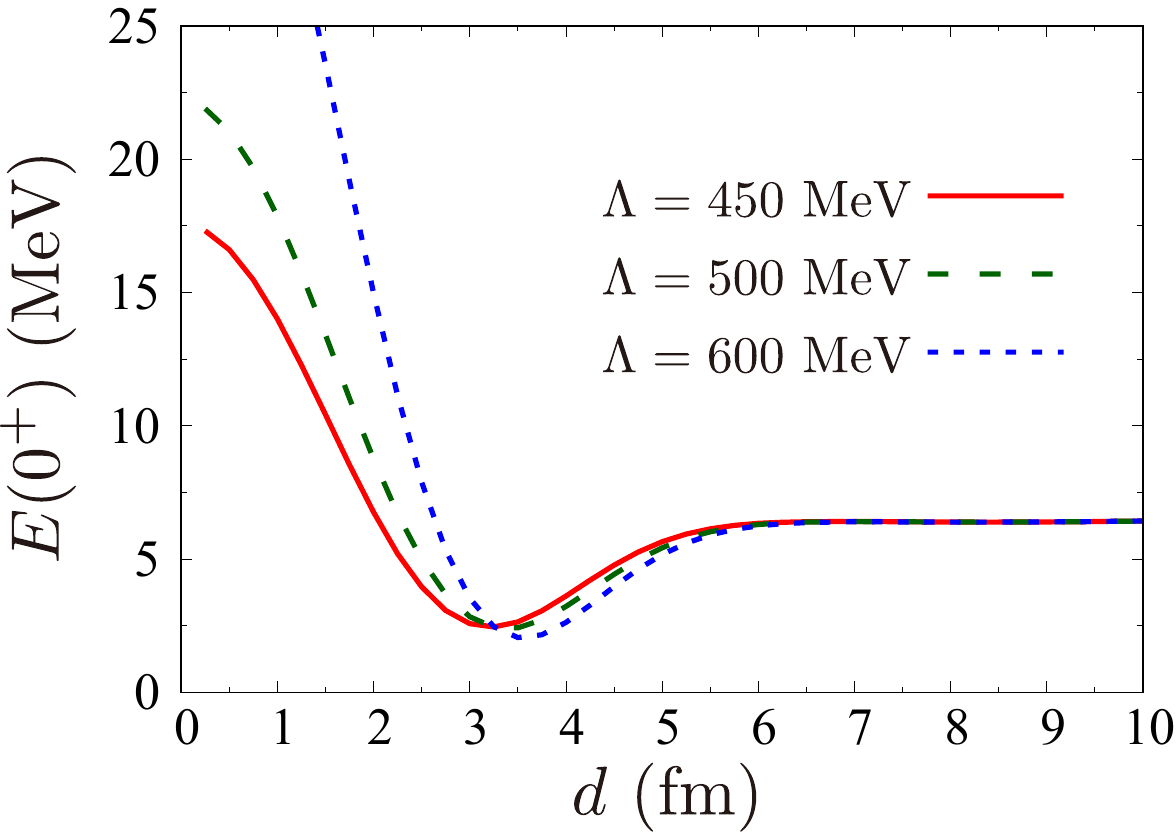}
 \caption{The relative energy $E(0^+)$ of $^8\mathrm{Be}$ measured from the $\alpha+\alpha$ threshold
 as a function of the relative distance between the clusters.
 The solid, dashed, and dotted lines are the results obtained with the cutoffs 
 $\Lambda=450,~500$, and $600$~MeV, respectively.}
\label{figE-d}
\end{center}
\end{figure}
In order to address the relative motion of the clusters, we define the relative energy $E(J^\pi)$ of $^8\mathrm{Be}$ 
measured from the $\alpha+\alpha$ threshold as
\begin{align}
 E(J^\pi)=E_{\mathrm{Be}}(J^\pi)-2E_\alpha,
 \label{EJpi}
\end{align}
where $E_{\mathrm{Be}}(J^\pi)$ is the eigenenergy of $\hat H_{\mathrm{eff}}$
associated with the total spin and parity $J^\pi$ state of $^8\mathrm{Be}$,
and $E_\alpha$ listed in Table~\ref{tableEalpha} are used.

\begin{figure}[!b]
\begin{center}
\includegraphics[width=0.70\textwidth,clip]{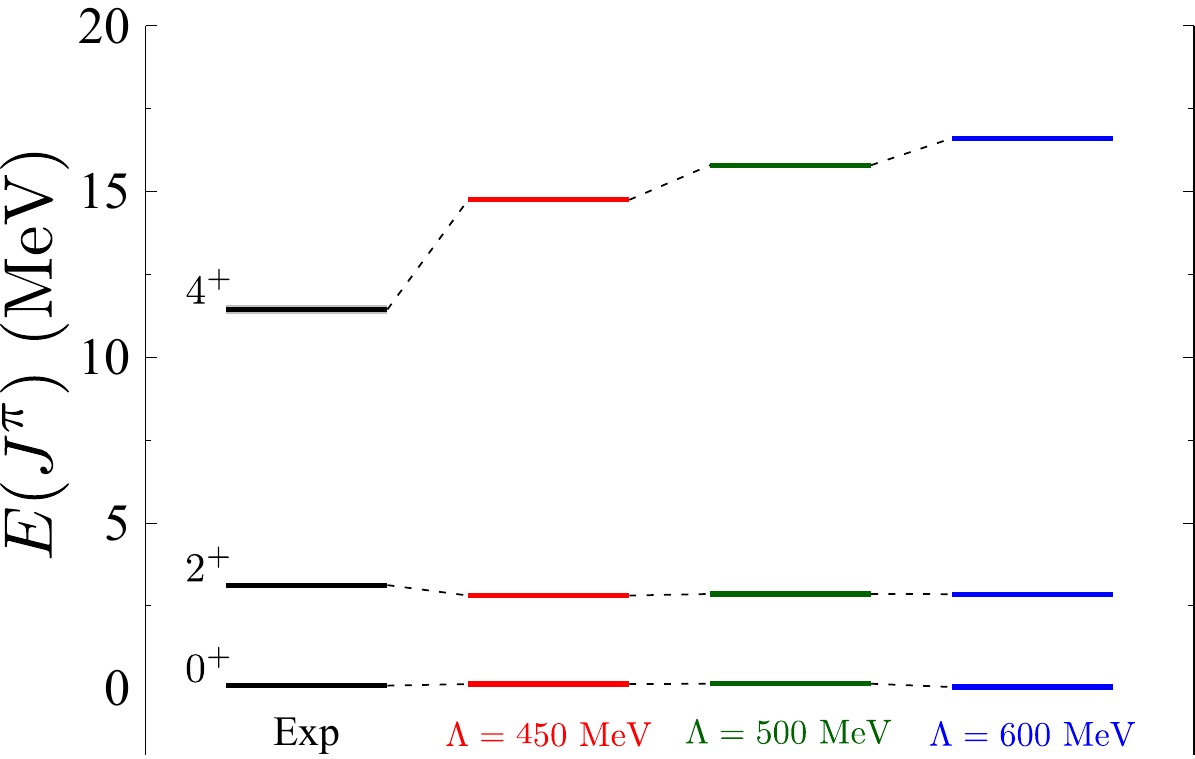}
 \caption{The low-lying spectra of $^8\mathrm{Be}$, referred to the $\alpha+\alpha$ threshold.
 The calculated levels denoted by $\Lambda=450,~500$, and $600$~MeV
 are compared with the observed data (Exp)~\cite{TILLEY2004155} in the left side.}
\label{figspectra}
\end{center}
\end{figure}
Figure~\ref{figE-d} shows the calculated results of $E(0^+)$ as a function of the $\alpha$-$\alpha$ distance $d$.
The solid, dashed, and dotted lines are obtained with $\Lambda=450,~500$, and $600$~MeV, respectively.
One finds that the energy minimum appears at $d\sim3.0$~fm with the 450-MeV cutoff and $d\sim3.5$~fm with the other cutoff values.
Since the 450-MeV-cutoff potential is less repulsive compared to that of the higher cutoffs, 
it describes the more compact ground state of $^8\mathrm{Be}$.
In Ref.~\cite{itagaki2020challenge}, it has demonstrated that, if the nucleon-nucleon potential is too repulsive, 
such energy minimum is not realized.
Thus, we find the current parameterization of $\hat H_{\mathrm{eff}}$ properly accounts for the attraction,
and the present results are qualitatively comparable to the results in Ref.~\cite{itagaki2020challenge},
where the phenomenological Volkov-No.2~\cite{VOLKOV196533} and Tohsaki-F1~\cite{PhysRevC.49.1814} interactions are employed.
The energy curves converge to a constant value at large $d$.
This is because $E(0^+)$ at large $d$, where the nucleon-nucleon interaction between the clusters vanishes,
can be estimated by the center-of-mass kinetic energy, $\hbar\omega/4\sim5$~MeV,
in addition to the small Coulomb energy.

\begin{figure}[!b]
\begin{center}
\includegraphics[width=1.0\textwidth,clip]{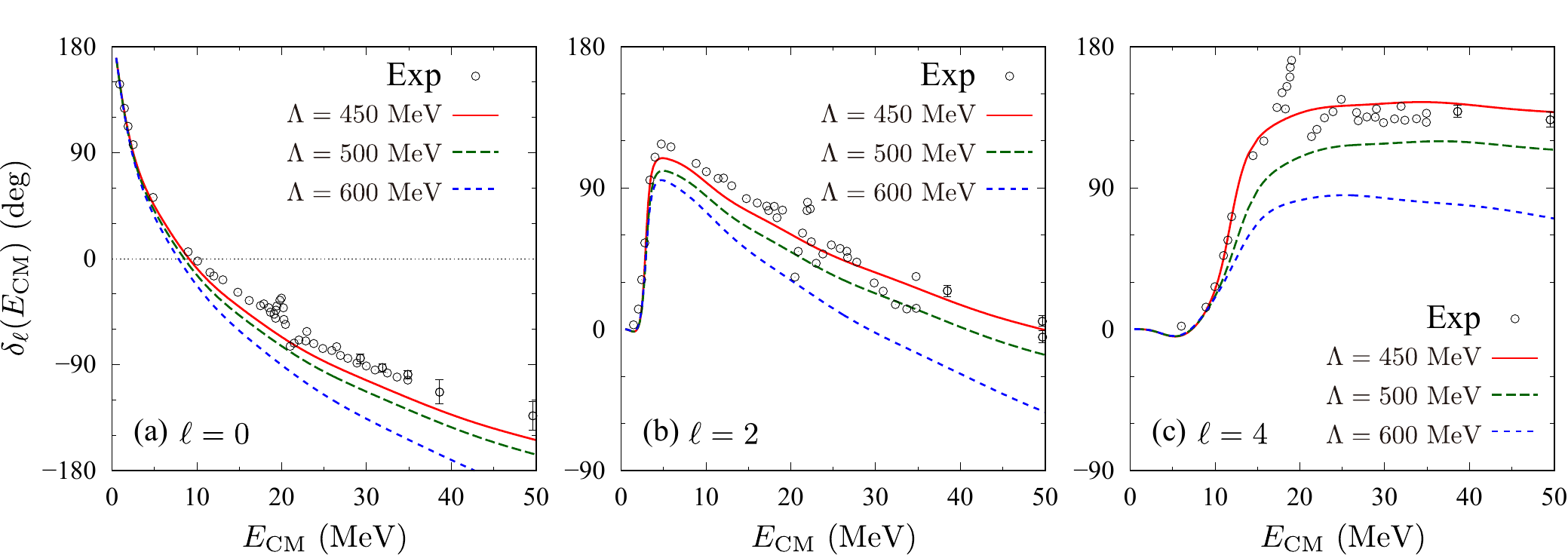}
 \caption{The scattering phase shift of the $\alpha+\alpha$ system of (a)~$\ell=0$, (b)~$\ell=2$, and (c)~$\ell=4$,
 as a function of the center-of-mass energy. 
 The solid, dashed, and dotted lines, respectively corresponding to the calculated results with the cutoffs 
 $\Lambda=450,~500$, and $600$~MeV,
 are confronted with the measured data~\cite{PhysRev.104.123,PhysRev.104.135,PhysRev.109.850,PhysRev.117.525,MiyakeBICR1961,PhysRev.129.2252,PhysRevLett.29.1331} represented by the circles.}
\label{figps}
\end{center}
\end{figure}

In Fig.~\ref{figspectra}, we compare the low-lying energy spectra of $^8\mathrm{Be}$
computed by the GCM with the experimental values~\cite{TILLEY2004155} displayed in the left side of the figure.
Tuning $c_0$, we obtain the $0^+$ energies (0.135, 0.146, and 0.050~MeV, respectively for 
$\Lambda=450,~500$, and $600$~MeV)
that coincide well with the measured energy,
and the $2^+$ energies are also consistent with the experimental one.
The calculations for the $4^+$ state overestimate the measured data by a few MeV
and become worse for the higher cutoffs.
This is probably due to the failure of the bound-state approximation of the present model,
which cannot describe properly the $4^+$ state as the broad resonance.
Indeed, as mentioned below, the scattering phase shift of the $\ell=4$ state are reasonably depicted,
in particular, with $\Lambda=450$ MeV (See Fig.~\ref{figps}).
We obtain the RMS-matter radius $R_\mathrm{M}$ of $^8\mathrm{Be}(0^+)$ calculated by the GCM
as 2.71, 2.76, and 2.79~fm associated with $\Lambda=450,~500$, and $600$~MeV, respectively.

Figure~\ref{figps} shows the scattering phase shift $\delta_{\ell}$ with the orbital-angular momentum $\ell$ 
associated with the relative motion of the $\alpha$+$\alpha$ system,
as a function of the center-of-mass energy $E_{\mathrm{CM}}$ between the clusters up to 50~MeV.
The solid, dashed, and dotted curves are the theoretical results corresponding to 
$\Lambda=450,~500$, and $600$~MeV, respectively.
The experimental $\delta_{\ell}$ represented by the circles are taken from Ref.~\cite{10.1143/PTP.53.677},
in which the original data were compiled from 
Refs.~\cite{PhysRev.104.123,PhysRev.104.135,PhysRev.109.850,PhysRev.117.525,MiyakeBICR1961,PhysRev.129.2252,PhysRevLett.29.1331}.
In Fig.~\ref{figps}(a), one finds 
that the three curves slightly underestimate the data of $\delta_{0}$,
whereas the underestimation by the 500-MeV and 600-MeV cutoffs is much more visible 
for the $\ell=2$ and $4$ cases exhibited in Figs.~\ref{figps}(b) and~\ref{figps}(c), respectively,
indicating the lack of the attraction.
The 500-MeV- and 600-MeV-cutoff potentials have 
more contributions from high momentum components compared to the 450-MeV-cutoff case.
In addition, these higher-cutoff potentials have 
relatively small components of the momentum less than the cutoff
because of the moderate regulator with small $n$ (see Table~\ref{tableconst2}).
These characteristics may lead to better results for $\Lambda=450$~MeV.

Our results on the phase shift are basically consistent with those in Ref.~\cite{10.1143/PTP.53.677},
where the data were reasonably explained by employing a nucleon-nucleon potential of Hasegawa and Nagata~\cite{10.1143/PTP.45.1786}.
This potential was designed to account for the binding energy of $^4\mathrm{He}$ and the $\alpha$-$\alpha$ scattering
phase shift by tuning the even-state attraction of the Tamagaki-A realistic potential~\cite{10.1143/PTPS.E68.242}.

\begin{figure}[!t]
\begin{center}
\includegraphics[width=0.70\textwidth,clip]{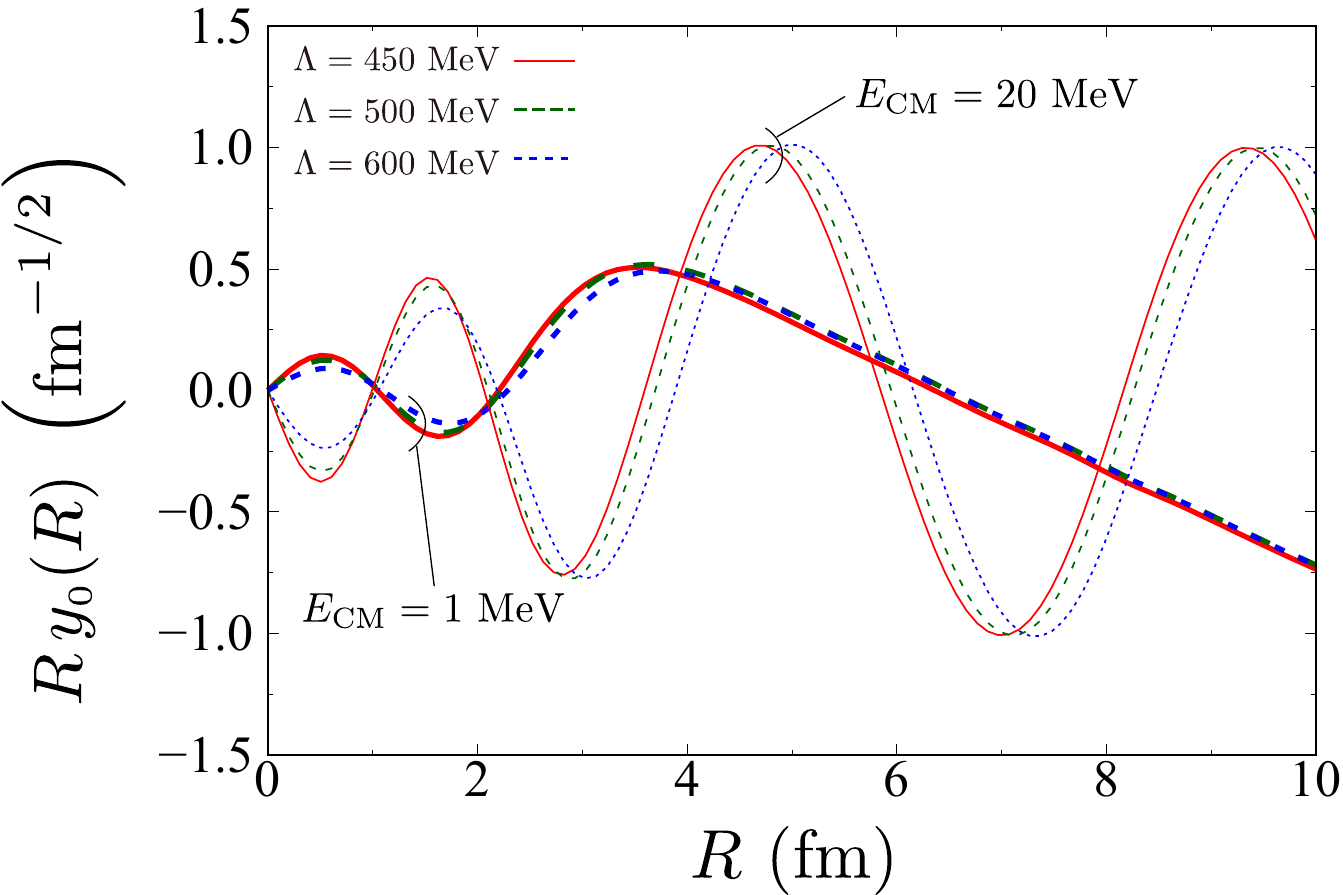}
 \caption{The reduced-width amplitude of the scattering state 
 with the orbital-angular momentum $\ell=0$ as a function of the relative-cluster distance.
 The solid, dashed, and dotted lines are obtained using 
 $\Lambda=450,~500$, and $600$~MeV, respectively.
 The thick lines (thin lines) correspond to $E_{\mathrm{CM}}=1$~MeV (20~MeV).}
\label{figRWA}
\end{center}
\end{figure}
Let us focus on the ground state of $^8\mathrm{Be}$.
Even though the $0^+$ energies computed with the three values of $\Lambda$ 
are almost equivalent with each other as shown in Fig.~\ref{figspectra},
the phase shift $\delta_0$ is significantly sensitive to the detail of the Hamiltonian.
Indeed, the $\Lambda$ dependence of the scattering states can be clearly seen in the reduced-width amplitude (RWA),
$y_{\ell}$, expressing the $\alpha$-$\alpha$ relative wave function.
The detail of calculating $y_{\ell}$ can be found in Refs.~\cite{10.1143/PTPS.62.11,10.1143/PTPS.62.90},
and its explicit form is also given in Appendix of Ref.~\cite{PhysRevC.93.034606}.
Figure~\ref{figRWA} displays the RWA of $\ell=0$ multiplied by the relative-cluster coordinate $R$, as a function of $R$.
The solid, dashed, and dotted lines correspond to the results with $\Lambda=450,~500$, and $600$~MeV, respectively.
One finds that, at $E_{\mathrm{CM}}=1$~MeV, 
where the phase shifts associated with the different $\Lambda$ are consistent with each other,
the RWAs expressed by the thick lines do not strongly depend on $\Lambda$.
In contrast, at $E_{\mathrm{CM}}=20$~MeV, the deviation of the phase of $y_{\scriptscriptstyle{0}}$ is visible among the thin lines,
in accordance with the difference appearing in the phase shift.
The similar correspondence between the phase shift and RWA is confirmed for the $\ell=2$ and $4$ states,
although the $\Lambda$ dependence of the RWA becomes drastic in these excited states.

\subsection{Possible origin of attraction}
\label{ResDis2}
\begin{figure}[!b]
\begin{center}
\includegraphics[width=1.0\textwidth,clip]{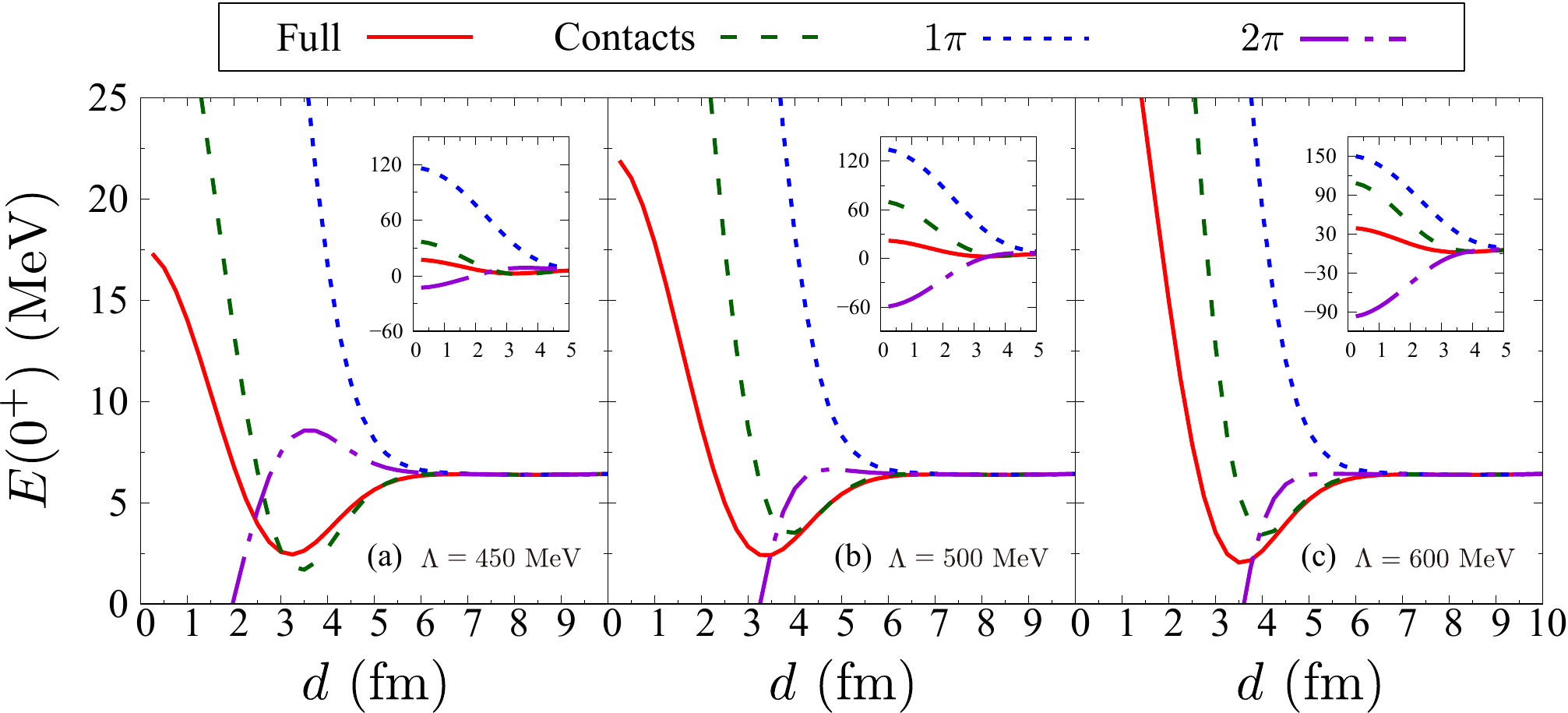}
 \caption{The relative energy of the $^8\mathrm{Be}$-ground state measured from the $\alpha+\alpha$ threshold
 computed with the chiral interaction at N$^3$LO (solid lines), the contact terms (dashed line),
 the $1\pi$-exchange term (dotted lines), and the $2\pi$-exchange terms (dash-dotted lines).
 The results of the cutoffs $\Lambda=450,~500$, and $600$~MeV 
 are shown in the left, middle, and right panels, respectively.}
\label{figE-d_indiv}
\end{center}
\end{figure}

Since the direct $1\pi$ exchange between nucleons inside the different $\alpha$ particles is forbidden,
sources of the attraction in the $\alpha$-cluster states need to be verified.
In order to make the discussion clear, we put our focus on the ground state of $^8\mathrm{Be}$.
We show $E(0^+)$ as a function of $d$ in Fig.~\ref{figE-d_indiv},
where the solid, dashed, dotted, and dash-dotted lines are obtained with 
the full contributions $\hat V_{ij}^{(\mathrm{N})}$ defined by Eq.~\eqref{chiralV},
the contacts $\hat V_{\mathrm{ct}}^{(0)}+\hat V_{\mathrm{ct}}^{(2)}+\hat V_{\mathrm{ct}}^{(4)}$, 
the $1\pi$-exchange term $\hat V_{1\pi}^{(0)}$, and the $2\pi$-exchange terms 
$\hat V_{2\pi}^{(2)}+\hat V_{2\pi}^{(3)}+\hat V_{2\pi}^{(4)}$, respectively.
The corresponding values of $E_\alpha$ are listed in Table~\ref{tableEalphaindiv}.
We employ $c_0$ given in Table~\ref{tablec0}.
As expected, the $1\pi$-exchange interaction in the $\alpha+\alpha$ system is crucially repulsive,
expressed by the dotted lines.
The contact interactions corresponding to the dashed lines are repulsive at small $d$,
but they seem to act attractively for $\Lambda=500$ and 600~MeV (450~MeV) at $d\sim4.0$~fm ($\sim 3.5$~fm),
where the energy minimum is produced, being qualitatively consistent with the results of the full calculations.
How the separated nucleons interact with each other via the contact force is explained later.
The $2\pi$-exchange interactions are strongly attractive, 
and thus we confirm the assumption~\cite{10.1143/PTP.27.793} that the main source of the attraction stems from these interactions.
\begin{table}[!t]
 \caption{Contributions of each pion-exchange term to the ground-state energy of $^4\mathrm{He}$.
 The value of $c_0$ fixed from the full contribution is commonly used.}
 \label{tableEalphaindiv}
 \begin{center}
   \begin{tabular}{c|lD{.}{.}{4}D{.}{.}{4}D{.}{.}{4}}
    \hline
    \multirow{5}{*}{$E_\alpha$ (MeV)} &  & \multicolumn{1}{c}{$\Lambda=450$ MeV} & \multicolumn{1}{c}{$\Lambda=500$ MeV} & \multicolumn{1}{c}{$\Lambda=600$ MeV} \\
    \cline{3-5}
                                      & Full     & -33.0397 & -34.3761 & -58.6637  \\
                                      & Contacts & -22.5984 & -16.5547 & -33.4038  \\
                                      & $1\pi$   & 97.5687  & 112.8210 &  125.9402 \\
                                      & $2\pi$   & -9.3107  & -31.9431 & -52.5009  \\
    \hline
   \end{tabular}
 \end{center}
\end{table}

\begin{figure}[!b]
\begin{center}
\includegraphics[width=1.0\textwidth,clip]{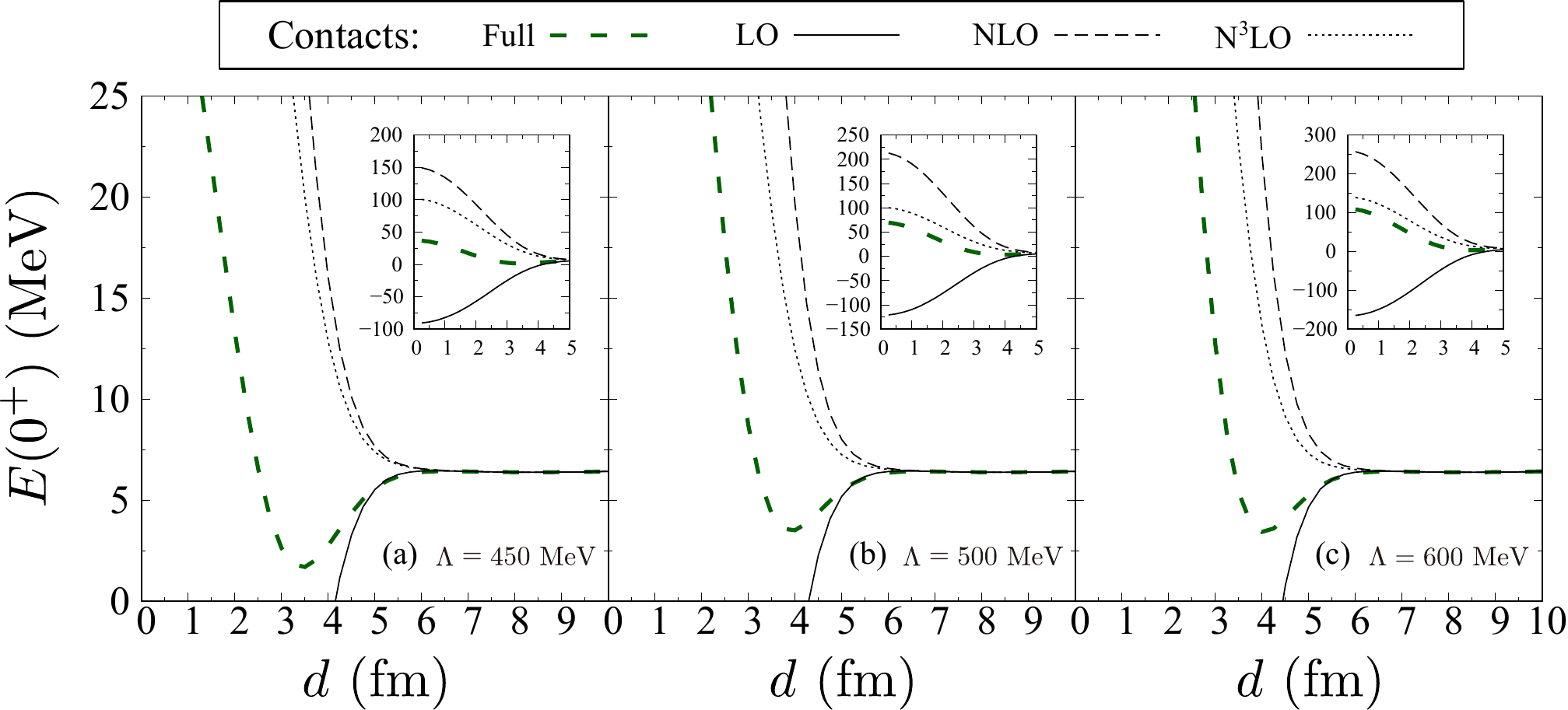}
 \caption{The contact-term contributions to $E(0^+)$.
 The thick-dashed, thin-solid, thin-dashed, and thin-dotted lines correspond to
 the results obtained with 
 $\hat V_{\mathrm{ct}}^{(0)}+\hat V_{\mathrm{ct}}^{(2)}+\hat V_{\mathrm{ct}}^{(4)}$,
 $\hat V_{\mathrm{ct}}^{(0)}$, $\hat V_{\mathrm{ct}}^{(2)}$, and $\hat V_{\mathrm{ct}}^{(4)}$,
 respectively, for the case of (a) $\Lambda=450$~MeV, (b) $\Lambda=500$~MeV, and (c) $\Lambda=600$~MeV.}
\label{figE-d_indiv_ct}
\end{center}
\end{figure}
\begin{table}[!t]
 \caption{A breakdown of the contact and $2\pi$ terms contributing to $E_\alpha$.}
 \label{tableEalphaindiv2}
 \begin{center}
   \begin{tabular}{c|lD{.}{.}{4}D{.}{.}{4}D{.}{.}{4}}
    \hline
     &  & \multicolumn{1}{c}{$\Lambda=450$ MeV} & \multicolumn{1}{c}{$\Lambda=500$ MeV} & \multicolumn{1}{c}{$\Lambda=600$ MeV} \\
    \cline{3-5}
     & \multicolumn{1}{l}{Contact at LO $\hat V_{\mathrm{ct}}^{(0)}$}      & -124.4714 & -155.1972 & -200.8181 \\
     & \multicolumn{1}{l}{Contact at NLO $\hat V_{\mathrm{ct}}^{(2)}$}     &  144.9496 &  187.2175 &  212.5949 \\
    \multirow{1}{*}{$E_\alpha$}
     & \multicolumn{1}{l}{Contact at N$^3$LO $\hat V_{\mathrm{ct}}^{(4)}$} &  55.6226  &   50.1244 &   53.5186 \\
    \multirow{1}{*}{(MeV)}
     & \multicolumn{1}{l}{$2\pi$ at NLO $\hat V_{2\pi}^{(2)}$}             &  59.8968  &   62.7673 &   66.0032 \\
     & \multicolumn{1}{l}{$2\pi$ at N$^2$LO $\hat V_{2\pi}^{(3)}$}         & -78.6510  & -127.9064 & -173.1798 \\
     & \multicolumn{1}{l}{$2\pi$ at N$^3$LO $\hat V_{2\pi}^{(4)}$}         & 108.1427  &  131.8953 &  153.3750 \\
    \hline
   \end{tabular}
 \end{center}
\end{table}
\begin{figure}[!b]
\begin{center}
\includegraphics[width=1.0\textwidth,clip]{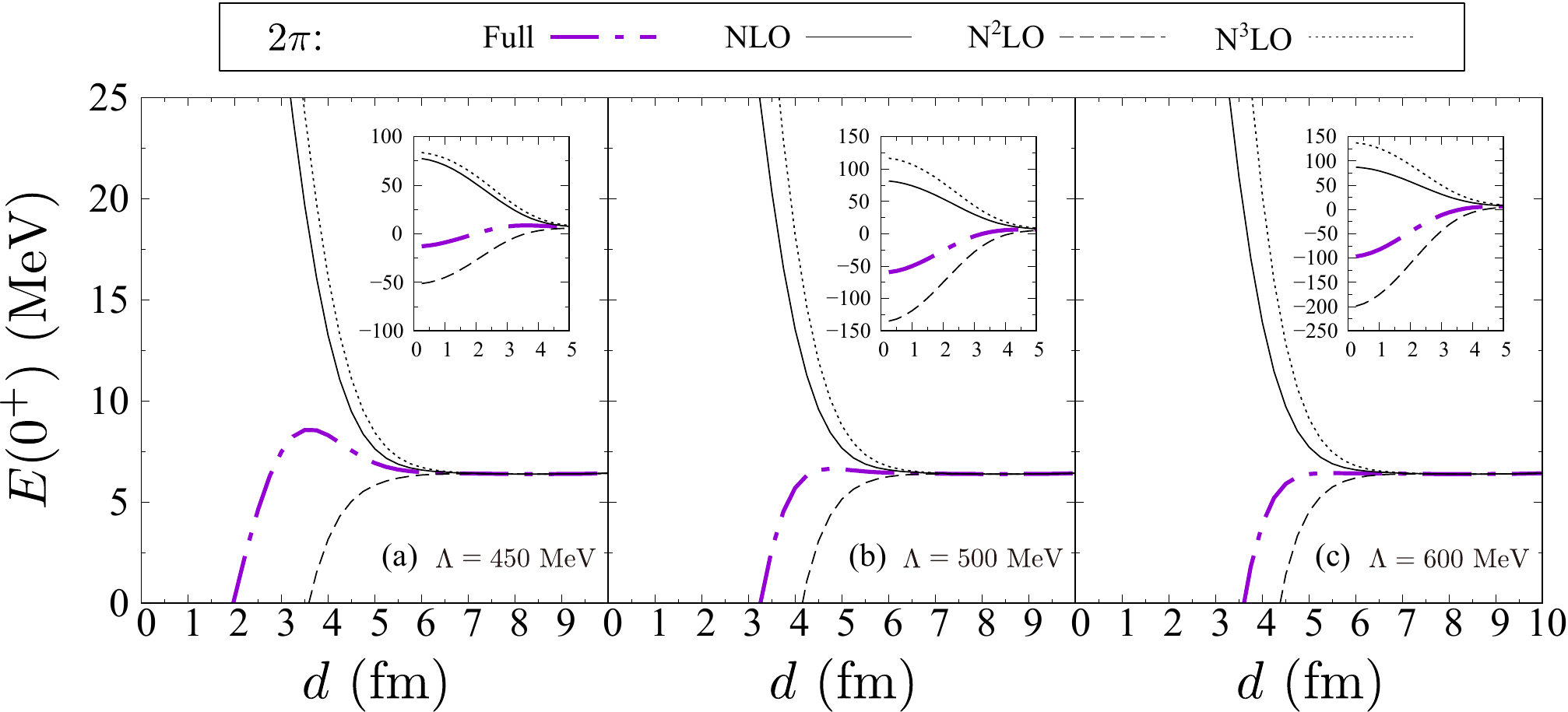}
 \caption{Same as Fig.~\ref{figE-d_indiv_ct}, but for the $2\pi$ contributions,
 and now the thick-dash-dotted, thin-solid, thin-dashed, and thin-dotted lines are 
 obtained by adopting 
 $\hat V_{2\pi}^{(2)}+\hat V_{2\pi}^{(3)}+\hat V_{2\pi}^{(4)}$,
 $\hat V_{2\pi}^{(2)}$, $\hat V_{2\pi}^{(3)}$, and $\hat V_{2\pi}^{(4)}$,
 respectively.}
 \label{figE-d_indiv_2pi}
\end{center}
\end{figure}
Here, we mention two points; 
(i) the attraction by the $2\pi$-exchange interactions alone is not sufficient 
to explain the ground-state property of $^8\mathrm{Be}$,
and (ii) the $2\pi$-exchange interactions of $\Lambda=450$~MeV behave quite differently from the others.
As regards the point (i), the short-range attraction originating from $\hat V_{\mathrm{ct}}^{(0)}$
plays also an essential role, consistent with the fact that 
the energy minimum is realized by the contact interactions alone in Fig.~\ref{figE-d_indiv}.
The attraction by the LO-contact term is demonstrated in Fig.~\ref{figE-d_indiv_ct}, 
where the thick-dashed, thin-solid, thin-dashed, and thin-dotted lines are $E(0^+)$
computed with $\hat V_{\mathrm{ct}}^{(0)}+\hat V_{\mathrm{ct}}^{(2)}+\hat V_{\mathrm{ct}}^{(4)}$ 
(equivalent to the dashed line in Fig.~\ref{figE-d_indiv}),
$\hat V_{\mathrm{ct}}^{(0)}$, $\hat V_{\mathrm{ct}}^{(2)}$, and $\hat V_{\mathrm{ct}}^{(4)}$, 
respectively. 
The values of $E_\alpha$ computed with these terms are given in Table~\ref{tableEalphaindiv2}.
In Fig.~\ref{figE-d_indiv_ct}, one finds that the behavior of energies are qualitatively independent of $\Lambda$.

\begin{figure}[!b]
\begin{center}
\includegraphics[width=1.0\textwidth,clip]{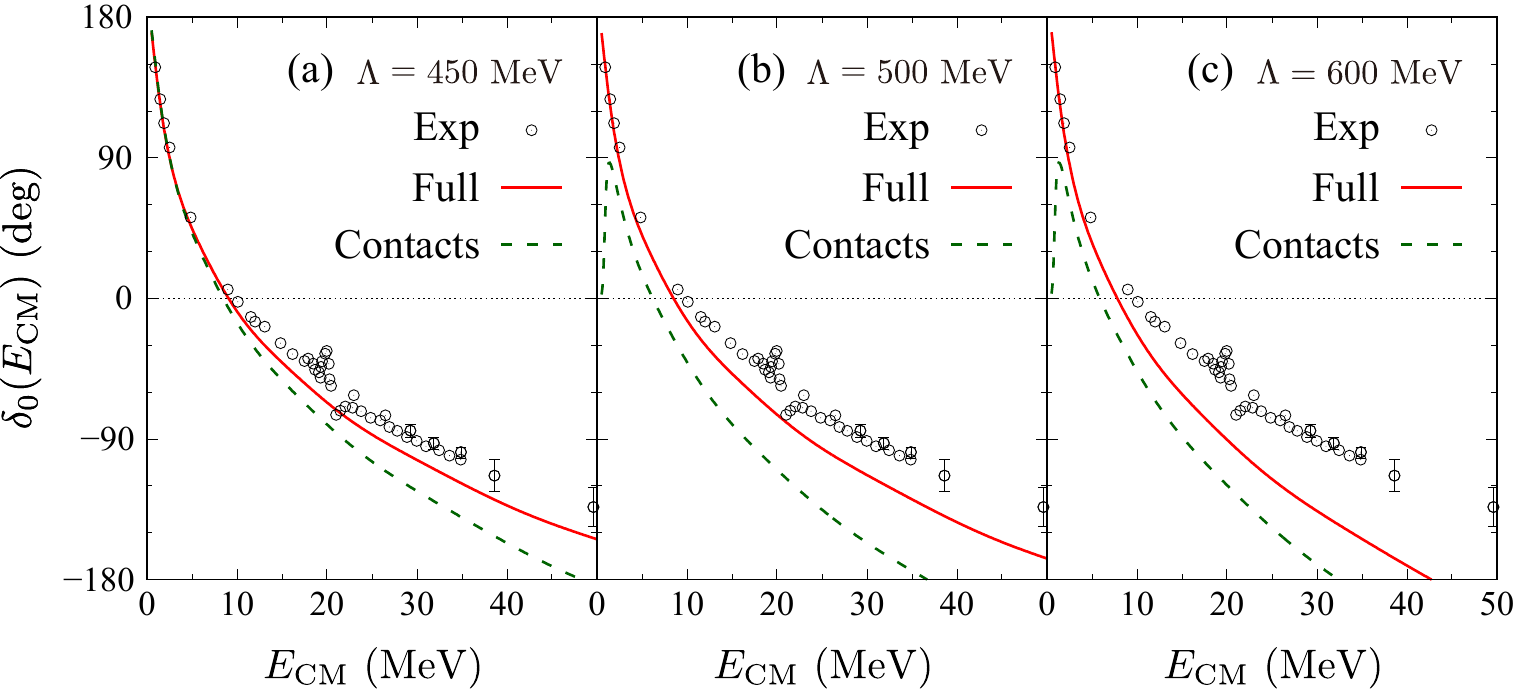}
 \caption{The phase shift of $\alpha$+$\alpha$ scattering with $\ell=0$ computed with the chiral interaction at N$^3$LO (solid lines)
 and the contact terms alone (dotted lines)
 for (a) $\Lambda=450$~MeV, (b) $\Lambda=500$~MeV, and (c) $\Lambda=600$~MeV. }
\label{figps_indiv}
\end{center}
\end{figure}
The point (ii) is due to the different constituents of the $2\pi$-exchange terms,
namely the N$^2$LO terms $\hat V_{2\pi}^{(3)}$ and the N$^3$LO terms $\hat V_{2\pi}^{(4)}$,
since the LECs $c_i$ of $\Lambda=450$~MeV are different from those of higher $\Lambda$ (see Table~\ref{tableconst2}).
To see this clearly, we show in Fig.~\ref{figE-d_indiv_2pi} 
a breakdown of the $2\pi$ contributions to $E(0^+)$,
computed with $\hat V_{2\pi}^{(2)}+\hat V_{2\pi}^{(3)}+\hat V_{2\pi}^{(4)}$,
$\hat V_{2\pi}^{(2)}$, $\hat V_{2\pi}^{(3)}$, and $\hat V_{2\pi}^{(4)}$,
respectively represented by the thick-dash-dotted, thin-solid, thin-dashed, and thin-dotted lines.
The corresponding calculations for $^4\mathrm{He}$ result in $E_\alpha$ 
given in Table~\ref{tableEalphaindiv2}.
One finds that, independently of $\Lambda$, $\hat V_{2\pi}^{(3)}$ is attractive
accounting for the intermediate-range attraction of the nuclear force 
as we know~\cite{MACHLEIDT20111},
while $\hat V_{2\pi}^{(4)}$ acts repulsively as comparable as $\hat V_{1\pi}^{(0)}$.
With decreasing $\Lambda$, the repulsion by $\hat V_{2\pi}^{(4)}$ 
becomes modest, but the attraction by $\hat V_{2\pi}^{(3)}$ is also suppressed, 
resulting in the less-attractive interaction totally of the $2\pi$-exchange contributions
$\hat V_{2\pi}^{(2)}+\hat V_{2\pi}^{(3)}+\hat V_{2\pi}^{(4)}$.
The contact terms at LO, $\hat V_{\mathrm{ct}}^{(0)}$, 
is another source of the attraction as shown in Fig.~\ref{figE-d_indiv_ct}, 
but it is not enough to compensate a loss of attraction in $\hat V_{2\pi}^{(3)}$ of $\Lambda=450$~MeV.
The full results are obtained by summing up all individual contributions coherently,
and thus, the interference through such a coherent process brings about
the whole interaction $\hat V_{ij}^{(\mathrm{N})}$
showing relatively weak $\Lambda$ dependence.

Since $E(0^+)$ computed with the contact interactions has the minimum point reasonably, 
we also investigate the behavior of the phase shift calculated with the contacts alone.
Figures~\ref{figps_indiv}(a), (b), and (c) show $\delta_0$ computed with $\Lambda=450,~500$, and $600$~MeV, respectively.
The thick lines are obtained by the full calculations including up to the N$^3$LO contributions,
while the dashed lines correspond to the results of the contact terms.
In Fig.~\ref{figps_indiv}(a), the two lines qualitatively agree with each other,
although the contact contributions are less attractive, as deduced also from Fig.~\ref{figE-d_indiv}(a).
The gross behavior of the dashed curves in Figs.~\ref{figps_indiv}(b) and (c) is not far from that of the solid line,
but we can find a distinguished difference at low $E_{\mathrm{CM}}$;
an unphysical peak appears around $E_{\mathrm{CM}} \sim 1.5$~MeV.
This is due to the poor reproduction of $E(0^+)$, which essentially determines the phase shift
at lower-scattering energies.
Indeed, with the contact terms alone, the GCM calculations result in
$E(0^+)=-0.0344, 1.1130$, and $1.1029$~MeV for $\Lambda=450,~500$, and $600$~MeV, respectively.

\begin{figure}[!b]
\begin{center}
 \includegraphics[width=0.70\textwidth,clip]{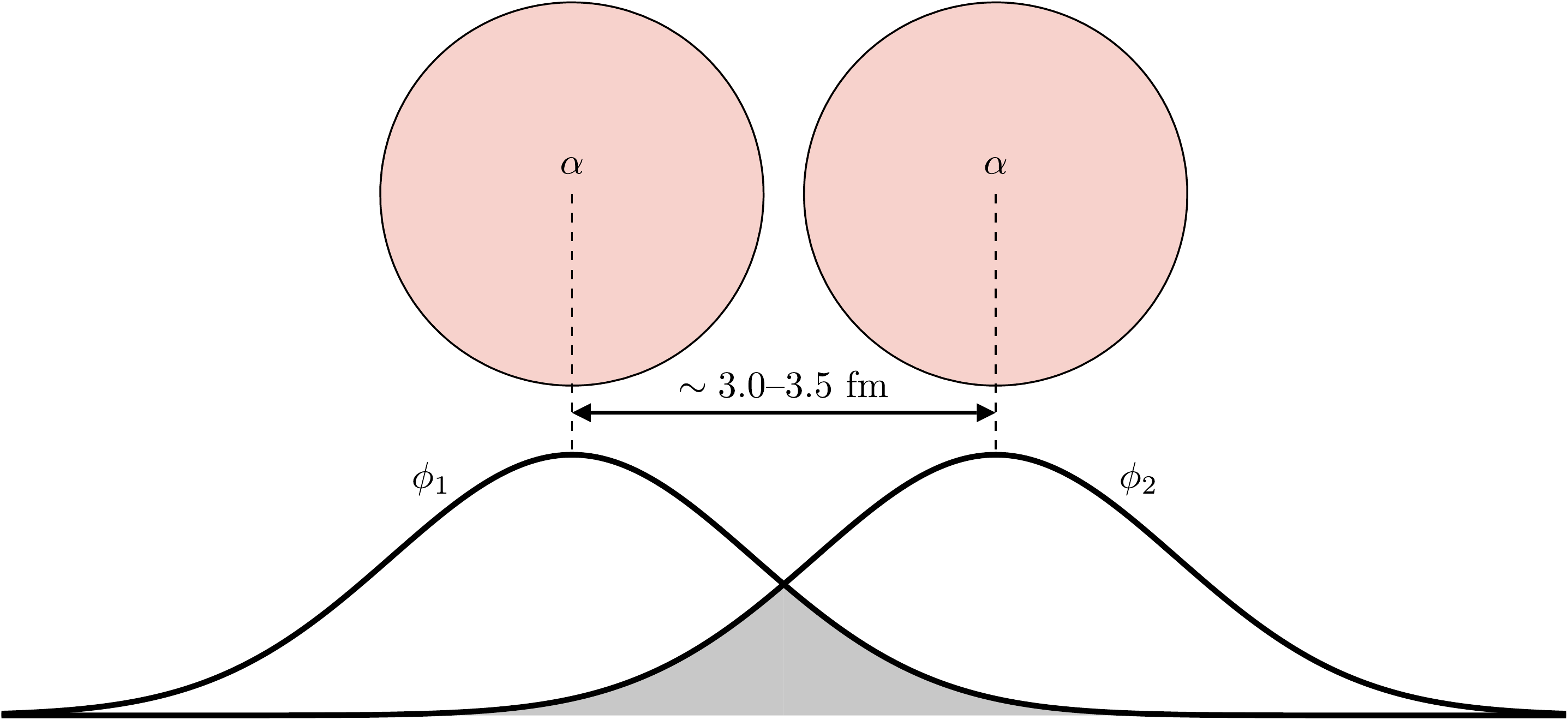}
 \caption{Intuitive picture explaining the mechanism how separated nucleons interact with each other via the contact force. 
 Two $\alpha$ particles are illustrated as the red circles,
 and the nucleon-wave packets, $\phi_1$ and $\phi_2$, have the overlap region expressed by the shadow.
 See the text for detail.}
\label{figintuitive}
\end{center}
\end{figure}
In conclusion, we can state that the attraction between nucleons in the $\alpha$-cluster states seems to originate dominantly from 
the $2\pi$-exchange terms, and the contact forces also plays an important role;
$\hat V_{2\pi}^{(3)}$ and $\hat V_{\mathrm{ct}}^{(0)}$ are attractive but the other terms are repulsive.
An intuitive picture to explain how the contact interactions contribute to a pair of separated nucleons
in the cluster states of $^8\mathrm{Be}$ is shown in Fig.~\ref{figintuitive}.
The two clusters are located having the relative distance $\sim 3.0$--3.5~fm,
as illustrated by the red circles.
The Brink model we adopt describes the nucleon distribution by the Gaussian-wave packets, 
the range of which is about 2~fm, corresponding to $\nu=0.26$~fm$^{-2}$.
The two wave packets, $\phi_1$ and $\phi_2$, which are settled at the center of each cluster,
overlap as expressed by the shadow.
Thus, we find that, even if the two clusters are separated from each other, 
nucleons can interact through the contact forces in the overlap region,
besides the main contribution of the attraction by the $2\pi$-exchange interactions.

In Ref.~\cite{10.1143/PTP.27.793}, where the $2\pi$-exchange interaction was assumed to be a source of the attraction,
the authors did not consider the short-range attraction.
Instead, they used the short-ranged-repulsive interaction expressed by a Gaussian function of the range $\sim 0.4$~fm.
Although their assumption is consistent with our outcome, 
such modeling of the nuclear force seems to result in a picture different from that of the chiral EFT.
In Ref.~\cite{PhysRevLett.117.132501}, the chiral interactions at only LO are included
in the nuclear-lattice EFT, by which the simulations of the eight-nucleon system reasonably explained 
the $\alpha$-$\alpha$ phase shift in the low-energy region.
The contact-origin attraction they adopted is not in contradiction with our consequence.

At the end of this section, we mention the relationship between the chiral EFT and
the conventional meson theory of nuclear forces~\cite{Machleidt1989,PhysRevC.63.024001},
since it can support our results.
Both theories provide high-precision nuclear potentials, which are quantitatively equivalent,
and hence, there is the correspondence between them.
The $1\pi$-exchange contribution is common but the other contributions are not bijection between both theories.
Roughly speaking, the $\sigma$-exchange contribution in the meson theory 
responsible for the intermediate-range attraction is fragmented into the $2\pi$-exchange terms order by order in the chiral EFT,
indicating that $2\pi$-exchange contributions $\hat V_{2\pi}^{(2)}+\hat V_{2\pi}^{(3)}+\hat V_{2\pi}^{(4)}$
totally would be attractive.
Similarly, the $\omega$-exchange contribution characterizing the repulsive core at short distances
corresponds naively to the chiral-contact terms order by order~\cite{MACHLEIDT20111},
and thus we expect $\hat V_{\mathrm{ct}}^{(0)}+V_{\mathrm{ct}}^{(2)}+\hat V_{\mathrm{ct}}^{(4)}$
would be repulsive at short distances.
These correspondences are consistent with our results,
although, as mentioned above, they are not bijection completely.
Hence, one of the $2\pi$-exchange (contact) contributions can be repulsive (attractive),
as confirmed by our calculations.

\section{Summary and perspectives}
\label{Summary}
With the aim of searching the possible source of the attraction
in the cluster states of $^8\mathrm{Be}$, we have phenomenologically prepared 
the effective interaction originating from the chiral EFT,
which can classify the pion-exchange contributions order by order.
Although we have introduced the phenomenological prefactor into the chiral potential,
the LECs are unchanged, and thus the relative contributions of each term of the chiral interaction
remain realistic.

Adopting the Brink model, the relative energy of $^8\mathrm{Be}$ measured from the two-$\alpha$ threshold
has been calculated and it has found to have a minimum point around 3.0 to 3.5~fm 
of the $\alpha$-$\alpha$ relative distance.
Performing the GCM calculations with respect to the relative-cluster distance as the generator coordinate,
we have computed the low-lying energy spectra and the scattering phase shift,
which are satisfactorily comparable to the experimental data.

We have investigated the individual contributions of the pion-exchange interactions at each chiral-expansion order,
and found that the $2\pi$-exchange terms can be a dominant source of the attraction
in the ground state of $^8\mathrm{Be}$, confirming the assumption of the previous work~\cite{10.1143/PTP.27.793}.
However, the short-range attraction originating from the contact terms in addition to the $2\pi$-exchange interactions
is also the important mechanism relevant to the $\alpha$-cluster states.

Our outcome in the present paper is not based on a fully microscopic approach, and therefore further investigation is necessary,
since (I) we have introduced a phenomenological prefactor, and (II) the three-nucleon force and the tensor-force effect are disregarded.
To overcome the point (I), we are now working on constructing cluster models based on the Brueckner theory
to microscopically drive the effective Hamiltonian starting from the chiral interaction.
It will be interesting to investigate whether the individual contributions of the chiral interaction remain unchanged 
in the effective interaction after solving the Bethe-Goldstone equation with the original force.
The formalism of the two-body MEs reported in the present article is applicable to such models.
In parallel with the construction of the realistic cluster model with the chiral interaction,
in order to tackle the point (II), we will take into account 
the chiral three-nucleon force at N$^2$LO~\cite{PhysRevC.49.2932,EpelbaumPhysRevC.66.064001,Epelbaum2006654,MACHLEIDT20111},
which is dominated by the $2\pi$-exchange terms, contributing to the tensor force.
As demonstrated in Ref.~\cite{PhysRevC.87.014327} the chiral three-nucleon force acts attractively 
in the $p$-shell nuclei including $^8\mathrm{Be}$, while the two-body tensor force provides repulsive effect~\cite{PhysRevC.98.054306,10.1093/ptep/ptz046}.
The interplay between them in the cluster states of $^8\mathrm{Be}$ needs to be clarified
to draw a robust conclusion on the origin of the attraction on such states.

\section*{Acknowledgment}
The author thanks N. Itagaki for helpful advice and providing a numerical code of the Brink model.
He also thanks M. Kamimura for supplying a numerical code to calculate the scattering phase shift,
and R. Machleidt for parameterizing the low-energy constants.
He is grateful to Y. Yamaguchi for fruitful discussions.
This work was supported in part by JSPS KAKENHI Grant Number JP21K13919.
The calculations have been carried out using the computer facilities at Yukawa Institute for Theoretical Physics, Kyoto University,
and Research Center for Nuclear Physics, Osaka University.

\appendix
\section{Investigation of parameter dependence}
\label{SecParaDep}
\subsection{Global prefactor}
\label{SecParaDepc0}
\begin{figure}[!b]
\begin{center}
\includegraphics[width=1.00\textwidth,clip]{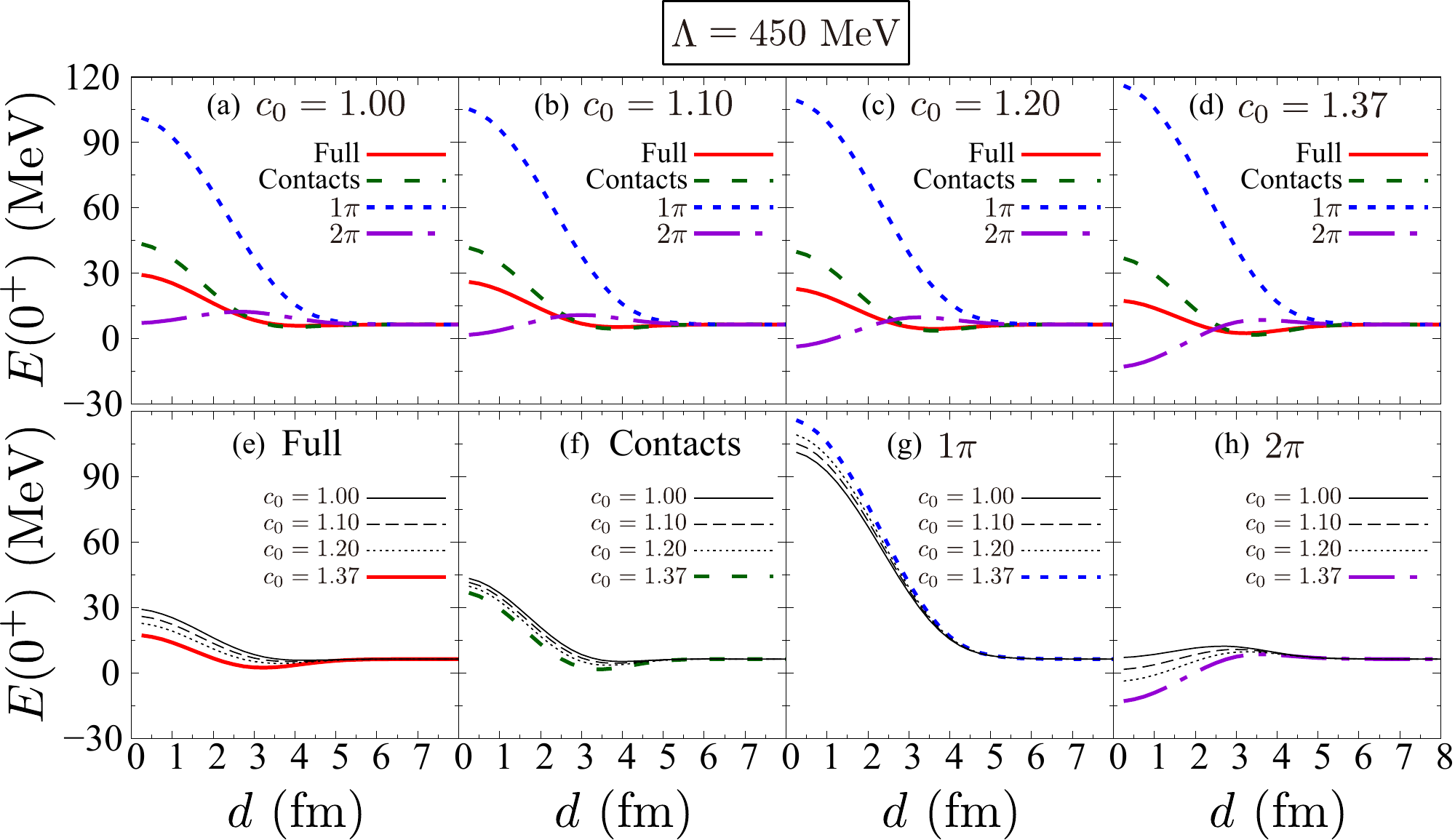}
 \caption{The $c_0$ dependence of $E(0^+)$ of $^8\mathrm{Be}$ for the $\Lambda=450$ MeV case.
 (a)--(d) The solid, dashed, dotted, and dash-dotted lines are obtained with the full, contact, $1\pi$, and $2\pi$
 contributions, respectively.
 (e)--(g) Thick lines correspond to the results of $c_0=1.37$, 
 while thin-solid, thin-dashed, and thin-dotted lines are those of $c_0=1.00$, $1.10$, and $1.20$, respectively.
 See text for details.}
\label{figE-d_evol}
\end{center}
\end{figure}
Here, we discuss how the results depend on the variation of $c_0$ from unity to the value 
fixed from the relative energy of the $^8\mathrm{Be}$-ground state, listed in Table~\ref{tablec0}.
Figure~\ref{figE-d_evol} displays $E(0^+)$ as a function of $d$
computed with each term of the interaction of $\Lambda=450$ MeV.
In Figs.~\ref{figE-d_evol}(a)--(d), where the legends are same as those in Fig.~\ref{figE-d_indiv},
$c_0$ is varied gradually from 1.00 to 1.37.
Independently of $c_0$, we can draw the conclusion as in Sec.~\ref{ResDis2},
i.e., the $2\pi$-exchange interaction plays a dominant role for the attraction for the cluster state,
the $1\pi$-exchange interaction is strongly repulsive,
and the contact interaction acts attractively at $d\sim3.5$ but basically repulsive at the short distances.
Note that the corresponding energies of $^{4}\mathrm{He}$ is listed in Table~\ref{tableEalpha_evol}.

Figures~\ref{figE-d_evol}(e), (f), (g), and (h) represent the $c_0$ dependence of $E(0^+)$ 
for the full contribution, contact terms, $1\pi$-exchange term, and $2\pi$-exchange terms.
The thick lines are the results of $c_0=1.37$, 
while those of $c_0=1.00$, $1.10$, and $1.20$ are the thin-solid, thin-dashed, and thin-dotted lines, respectively.
These lines are fragmented into Figs.~\ref{figE-d_evol}(a)--(d);
the thin-solid, thin-dashed, thin-dotted, and thick lines in Fig.~\ref{figE-d_evol}(e)
are equivalent to the solid lines in Figs.~\ref{figE-d_evol}(a), (b), (c), and (d), respectively,
and similar for the other lines in Figs.~\ref{figE-d_evol}(f)--(h).
One sees that the energies associated with each contribution vary gradually
with respect to the evolution of $c_0$.

We confirmed numerically that the similar results are obtained for $\Lambda=500$ and $600$ MeV,
even though the range of the $c_0$ variation is rather large compared to the 450-MeV-cutoff case.
Thus we confirm that our conclusion is robust against the variation of $c_0$.
\begin{table}[!t]
 \caption{The calculated ground-state energy of $^4\mathrm{He}$ with $\Lambda=450$~MeV
 when $c_0$ varies from unity to 1.37 fixed from the relative energy of the $^8\mathrm{Be}$-ground state.}
 \label{tableEalpha_evol}
 \begin{center}
   \begin{tabular*}{0.95\textwidth}{@{\extracolsep{\fill}}D{.}{.}{2}|cD{.}{.}{4}D{.}{.}{4}D{.}{.}{4}D{.}{.}{4}}
    \hline
    \multicolumn{1}{c|}{$c_0$}  &    & \multicolumn{1}{c}{Full} & \multicolumn{1}{c}{Contacts} & \multicolumn{1}{c}{$1\pi$} & \multicolumn{1}{c}{$2\pi$} \\
    1.00 &                                & -10.7886 &  -3.1672 & 84.5460 & 6.5319  \\
    1.10 & \multicolumn{1}{c}{$E_\alpha$} & -16.8024 &  -8.4189 & 88.0657 & 2.2501  \\
    1.20 & \multicolumn{1}{c}{(MeV)}      & -22.8162 & -13.6706 & 91.5853 & -2.0317 \\
    1.37 &                                & -33.0397 & -22.5984 & 97.5687 & -9.3107 \\
    \hline
   \end{tabular*}
 \end{center}
\end{table}

\subsection{Gaussian-range parameter}
\label{SecParaDepnu}
To investigate the $\nu$ dependence of $E(0^+)$, we perform the calculations with $\nu=0.24$, $0.26$, and $0.28~\mathrm{fm}^{-2}$.
When $\nu$ is changed, $c_0$ also needs to be modified slightly to describe the measured value of the relative energy of $^8\mathrm{Be}$.
For $\Lambda=450$~MeV, we obtain $E(0^+)$ computed by the GCM as 0.0453, 0.135, 0.122~MeV 
with $(\nu,c_0)=(0.24~\mathrm{fm}^{-2},1.36)$, $(0.26~\mathrm{fm}^{-2},1.37)$, and $(0.28~\mathrm{fm}^{-2},1.39)$, respectively.
\begin{figure}[!t]
\begin{center}
\includegraphics[width=0.70\textwidth,clip]{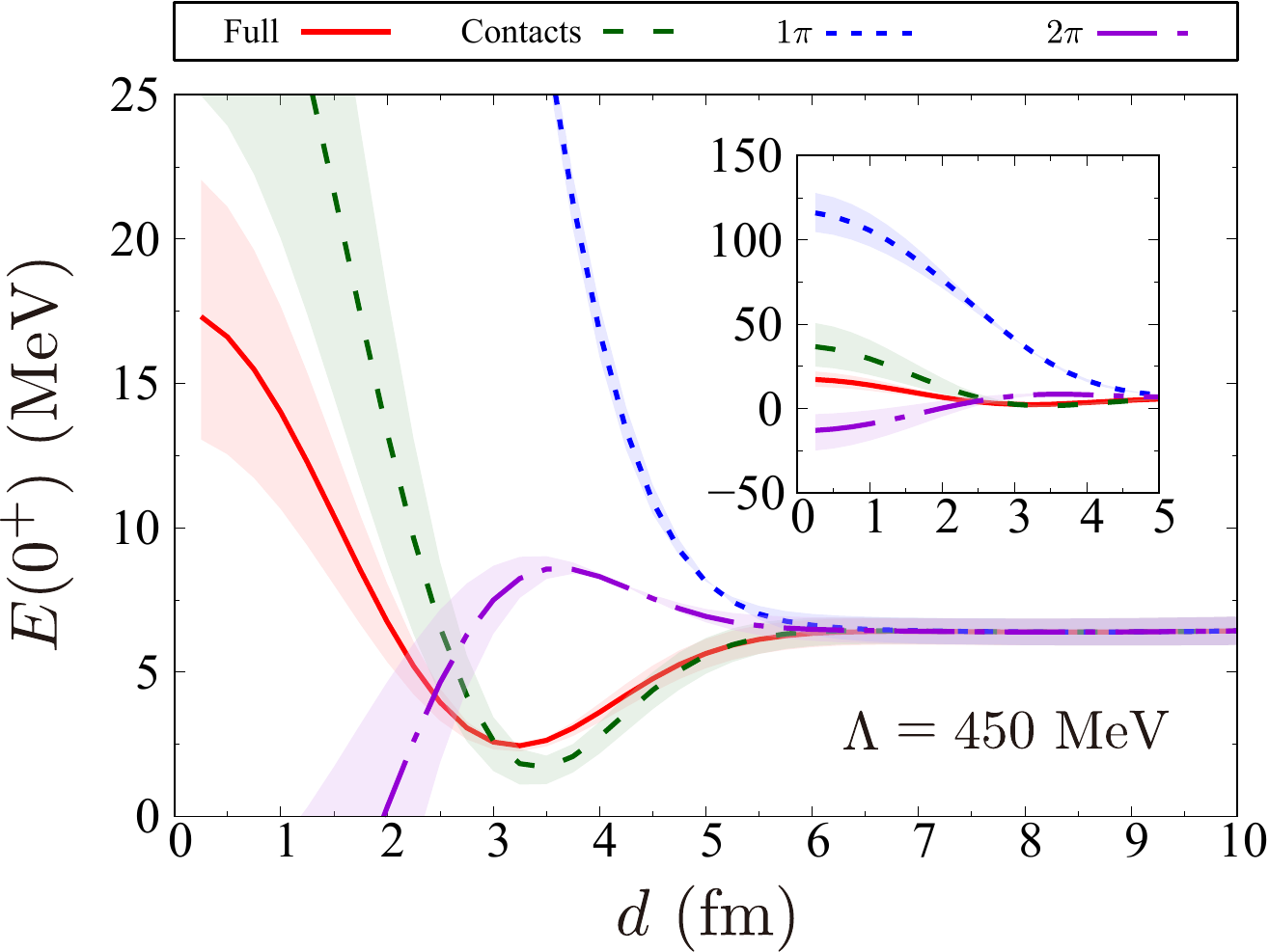}
 \caption{The $\nu$ dependence of $E(0^+)$ of $^8\mathrm{Be}$ for the $\Lambda=450$ MeV case.
 The solid, dashed, dotted, and dash-dotted lines are respectively the results for the full, contact, $1\pi$, and $2\pi$
 contributions with $\nu=0.26~\mathrm{fm}^{-2}$.
 The filled areas associated with each line are obtained when $\nu$ varies from $0.24$ to $0.28~\mathrm{fm}^{-2}$.
 See text for details.}
\label{figE-d_nu}
\end{center}
\end{figure}

Figure~\ref{figE-d_nu} shows $E(0^+)$ of $^8\mathrm{Be}$ as a function of $d$
computed with $\Lambda=450$~MeV.
The legends are same as those of Fig.~\ref{figE-d_indiv} but now the lines have the bands.
The lines correspond to the result with $\nu=0.26~\mathrm{fm}^{-2}$, as those in Fig.~\ref{figE-d_indiv}(a).
The calculations with $\nu=0.24~\mathrm{fm}^{-2}$ ($\nu=0.28~\mathrm{fm}^{-2}$) determine
the lower (upper) limits of the band of the full, contact, and $1\pi$ cases for small $d$,
while they oppositely results in the upper (lower) limits for the $2\pi$ case.
A breakdown of $E_\alpha$ computed with different $\nu$ is listed in Table~\ref{tableEalpha_nu}.

One finds from Fig.~\ref{figE-d_nu} that
our conclusion drawn in Sec.~\ref{ResDis2} remains unchanged
even when $\nu$ varies;
the $2\pi$-exchange contribution is a dominant source of the attraction,
the $1\pi$-exchange one gives strong repulsion,
and the contacts bring about the nonnegligible attraction at the intermediate $\alpha$-$\alpha$ distances
with mainly repulsive nature.
Similar numerical calculations confirmed that the variation of $\nu$ does not affect our conclusion
even for the other $\Lambda$.

\begin{table}[!t]
 \caption{The ground-state energy of $^4\mathrm{He}$ calculated with $\Lambda=450$~MeV
 for different values of $\nu$.}
 \label{tableEalpha_nu}
 \begin{center}
   \begin{tabular*}{0.95\textwidth}{@{\extracolsep{\fill}}D{.}{.}{4}|cD{.}{.}{4}D{.}{.}{4}D{.}{.}{4}D{.}{.}{4}}
    \hline
    \multicolumn{1}{c|}{$\nu$ ($\mathrm{fm}^{-2}$)} &  & \multicolumn{1}{c}{Full} & \multicolumn{1}{c}{Contacts} & \multicolumn{1}{c}{$1\pi$} & \multicolumn{1}{c}{$2\pi$} \\
    0.24 & \multicolumn{1}{c}{$E_\alpha$} & -31.0406 & -24.9839 &  87.2218 &  -2.1090 \\
    0.26 & \multicolumn{1}{c}{(MeV)}      & -33.0397 & -22.5984 &  97.5687 &  -9.3107 \\
    0.28 &                                & -35.2337 & -19.2930 & 108.6771 & -18.3912 \\
    \hline
   \end{tabular*}
 \end{center}
\end{table}

\section{Formulation of two-body matrix elements}
\label{SecMEs}
\subsection{General expression of matrix elements}
\label{SecMEsGenral}
We derive the two-body MEs of Eq.~\eqref{chiralV} with the antisymmetrized two-body states defined by
\begin{align}
 \ket{ijab}_A
 =
 \frac{1}{\sqrt 2} \left[\ket{\phi_i \chi_a }\ket{\phi_j \chi_b} -\ket{\phi_j \chi_b}\ket{\phi_i \chi_a }\right].
 \label{2bstates}
\end{align}
Owing to the spin and isospin saturation in the Brink model, 
only the central-force contributions of $\hat V_{ij}^{(\mathrm{N})}$ remain,
and thus all the antisymmetrized two-body MEs are reduced to the form,
\begin{align}
 &\sum_{a=1}^4\sum_{b=1}^4
 {}_{\substack{\vspace{4pt}\\A\!\!}}\left< klab \left| \left.
 \hat V_{\mathrm{X}}^{(n_\chi)} \right. \right| ijab \right>_A
 \nonumber\\
 &\quad=
 A_{ijkl}
 \sum_{\ell}(-)^{\ell}
 P_{\ell}(x_{ijkl})
 \nonumber\\
 &\quad\times
 \int\!\!\!\!\int
 dp dp'
 p^{2}p'^{2}
 g(p,p')  u_n\!\left(p,p'\right)
 j_{\ell}\!\left(r_{ij}p\right)  j_{\ell}\!\left(r_{kl}p'\right)
 \mathcal{M}_{\mathrm{X;\,}\ell}^{(n_\chi)}\!\left(p,p'\right).
 \label{generalME}
\end{align}
Here the symbol $\mathrm{X}$ is a representative for each pion-exchange contribution ($\mathrm{X}=\mathrm{ct}$, $1\pi$, or $2\pi$),
and $\mathcal{M}_{\mathrm{X;\,}\ell}^{(n_\chi)}$ is the multipole-expansion (MPE) function, the explicit form of which is given in \ref{SecMEsMPE}.
The center-of-mass contribution is factored out as $A_{ijkl}$ defined by
\begin{align}
 A_{ijkl}
 &=
 \frac{1}{\left(\pi\nu\right)^{\frac{3}{2}}}
 \exp\!\!\left[-\nu\left(\vect{R}_{ij}-\vect{R}_{kl}\right)^2\right],
 \label{Aijkl}\\
 \vect{R}_{ij} 
 &=
 \frac{1}{2}\left( \vect{R}_i +\vect{R}_j \right).
 \label{rCMij}
\end{align}
Two nucleons interact with each other characterized by the relative momentum $\vect{p}$ ($\vect{p}'$)
in the initial channel (final channel), associated with the orbital-angular momentum $\ell$.
We define the relative vector,
\begin{align}
 \vect{r}_{ij} &= \vect{R}_i -\vect{R}_j,
 \label{rrelij}
\end{align}
and its magnitude $r_{ij}$, which appears in the argument of the spherical Bessel function $j_{\ell}$,
is equivalent to the cluster distance $d$ introduced in Sec.~\ref{TheorFrameBrink}.
The Legendre polynomial $P_{\ell}$ has the argument $x_{ijkl}$ defined by
\begin{align}
 x_{ijkl}
 =
 \begin{cases}
  1~\quad & (i=j~\mathrm{and/or}~k=l),\\
  \cos\theta_{ijkl}
  =
  \frac{\vect{r}_{ij} \cdot \vect{r}_{kl}}{r_{ij}r_{kl}}
  ~\quad & (i\ne j~\mathrm{and}~k\ne l),
 \end{cases}
 \label{xijkl}
\end{align}
with the angle $\theta_{ijkl}$ between the vectors $\vect{r}_{ij}$ and $\vect{r}_{kl}$.
The function $g$ is given by
\begin{align}
 g(p,p')
 &=
 \exp\!\!\left[-\frac{1}{2\nu}\left(p^2+p'^2\right)\right],
 \label{gaussg}
\end{align}
while the regulator $u_n$ has the nonlocal form~\cite{MACHLEIDT20111} as
\begin{align}
 u_n\!\left(p,p'\right)
 &=
 \exp\!\left[-\!\left(\frac{p}{\Lambda}\right)^{\!\!2n}-\!\left(\frac{p'}{\Lambda}\right)^{\!\!2n}\right],
 \label{reg}
\end{align}
with the cutoff momentum $\Lambda$.

Equation~\eqref{generalME} is the expression relevant to the calculations before the angular-momentum projection.
To perform the angular-momentum projection by Eq.~\eqref{JMMstate},
we need to rotate the two-body states $\ket{ijab}_A$.
As a result, we just have to transform the vectors $\vect{R}_{ij}$ and $\vect{r}_{ij}$ as
\begin{align}
 \vect{R}_{ij} &\to \vect{R}_{ij}^{(R)}=\frac{1}{2}\left[ \hat{R}(\Omega)\vect{R}_i +\hat{R}(\Omega)\vect{R}_j \right],
 \label{rCMijRot}\\
 \vect{r}_{ij} &\to \vect{r}_{ij}^{(R)}=\hat{R}(\Omega)\vect{R}_i -\hat{R}(\Omega)\vect{R}_j.
 \label{rrelijRot}
\end{align}
The spin states are invariant under the spin-space rotation within the Brink model presuming the spin-singlet-nucleon pairs.
Furthermore, we do not need to rotate the vectors $\vect{R}_{kl}$ and $\vect{r}_{kl}$ associated with the bra states,
since the Hamiltonian we employ is rotationally invariant.

Note that it is necessary to multiply the conventional chiral potential~\cite{MACHLEIDT20111} by $1/(2\pi)^3$,
originating from our convention of the normalization, $\braket{\vect{p}|\vect{p}'}=\delta(\vect{p}-\vect{p}')$.
See Refs.~\cite{NavratilFS2007,PhysRevC.98.044305} for similar prefactors involved in the chiral three-nucleon potential.

\subsection{Multipole-expansion function}
\label{SecMEsMPE}
The potential of the chiral interaction defined in the momentum space depends on $\vect{p}$ and $\vect{p}'$~\cite{MACHLEIDT20111},
or equivalently the magnitude of the transferred momentum $\vect{q}$ and the average momentum $\vect{Q}$, respectively defined by
\begin{align}
 \vect{q}&=\vect{p}'-\vect{p},
 \label{transmom}\\
 \vect{Q}&=\frac{1}{2}\left(\vect{p}+\vect{p}'\right).
 \label{avemom}
\end{align}
We perform the MPE with respect to $x=\cos\theta$ with the angle $\theta$ between $\vect{p}$ and $\vect{p}'$.
Note that we do not need to perform the MPE for the contact terms,
while for the long-range terms, we need to do so. 
More explicitly, for the former, the angular
integration in the MPE function can be done analytically owing to
the absence of the pion propagator. However, for the latter,
the pion propagator prevents from performing the integration analytically,
and therefore numerical integration is necessary.

The LECs involved in the MPE functions defined below are explicitly given in \ref{SecConstants}.
Let us start to write the MPE function of the contact terms at LO as
\begin{align}
 \mathcal{M}_{\mathrm{ct;\,}\ell}^{(0)}\!\left(p,p'\right)
 &=
 \frac{4}{\pi}\left[C_S^{pp}+C_S^{nn}+4C_S^{nn}-3\left(C_T^{pp}+C_T^{nn}\right)\right]
 \delta_{\ell0},
 \label{MPELOct}
\end{align}
where the LECs $C_S$ and $C_T$ depend on the nucleon charge.
Obviously there is only the $s$-wave contribution here,
and it is natural since $\ell$ contributing to the pure contact terms at LO should be zero.

The next term is the $1\pi$-exchange contribution at LO, the MPE function of which reads,
\begin{align}
 \mathcal{M}_{1\pi;\,\ell}^{(0)}\!\left(p,p'\right)
 =
 \frac{6}{\pi}\left(\frac{g_A}{f_\pi}\right)^2
 \frac{\hat{\ell}^2}{2}\int_{-1}^{1} dx P_{\ell}(x)
 \frac{q^{2}}{q^2+m_\pi^2},
 \label{MPELO1pi}
\end{align}
where $g_A$, $f_\pi$, and $m_\pi$ are the axial vector coupling constant,
the pion-decay constant, and the average pion mass, respectively.
Their values, the average nucleon mass $m_N$ as well, are summarized in \ref{SecConstants}.
We adopt the abbreviation $\hat{\ell}=\sqrt{2\ell+1}$.

At NLO of the contact terms, the $p$ wave starts to contribute as
\begin{align}
 \mathcal{M}_{\mathrm{ct;\,}\ell}^{(2)}\!\left(p,p'\right)
 =
 &\frac{2}{\pi}\left[\left(12C_1+3C_2-12C_3-3C_4-4C_6-C_7\right)(p^2+p'^2)\delta_{\ell0}\right.
 \nonumber\\
 &+2\left.\left(20C_1-5C_2+12C_3-3C_4+4C_6-C_7\right)pp'\delta_{\ell1}\right],
 \label{MPENLOct}
\end{align}
where $C_i$ are the LECs.

The MPE function of the $2\pi$-exchange terms at NLO is expressed by
\begin{align}
 \mathcal{M}_{2\pi;\,\ell}^{(2)}\!\left(p,p'\right)
 &=
 \frac{1}{(2\pi)^3f_\pi^4}
 \frac{\hat{\ell}^2}{2}\int_{-1}^{1} dx P_{\ell}(x)
 \left[W_C^{(2)}(q,Q)+V_S^{(2)}(q,Q)\right],
 \label{MPENLO2pi}\\
 W_C^{(2)}(q,Q)
 &=
 \frac{1}{2}L(q)
 \left[4m_\pi^2\left(5g_A^4-4g_A^2-1\right)
 +q^2\left(23g_A^4-10g_A^2-1\right)
 +48g_A^4 m_\pi^4w^{-2}\right],
 \label{WCNLO}\\
 V_S^{(2)}(q,Q)
 &=
 -6g_A^4
 q^2L(q),
 \label{VSNLO}
\end{align}
and
\begin{align}
 L(q)
 &=
 \frac{w}{q}\ln\frac{w+q}{2m_\pi},
 \label{2piLq}\\
 w
 &=
 \sqrt{4m_\pi^2 +q^2}.
 \label{2piw}
\end{align}

At N$^2$LO we have
\begin{align}
 \mathcal{M}_{2\pi;\,\ell}^{(3)}\!\left(p,p'\right)
 &=
 \frac{3g_A^2}{(4\pi)^2f_\pi^4}
 \frac{\hat{\ell}^2}{2}\int_{-1}^{1} dx P_{\ell}(x)
 \nonumber\\
 &\times
 \left[\left\{4(-)^{\ell}-1\right\}V_C^{(3)}(q,Q)+W_C^{(3)}(q,Q)+V_S^{(3)}(q,Q)+W_S^{(3)}(q,Q)\right],
 \label{MPEN2LO2pi}\\
 V_C^{(3)}(q,Q)
 &=
 8\left[\frac{g_A^2 m_\pi^5}{16 m_N w^2}
 -\left\{2m_\pi^2\left(2c_1-c_3\right) -q^2\left(c_3 +\frac{3g_A^2}{16m_N}\right)\right\}
 \tilde w^2 A(q)
 \right.
 \nonumber\\
 &\left.
 -\frac{g_A^2}{16 m_N}\left\{m_\pi w^2 +\tilde w^4 A(q)\right\}\right],
 \label{VCN2LO}\\
 W_C^{(3)}(q,Q)
 &=
 -\frac{1}{m_N}
 \left[
 3g_A^2 m_\pi^5 w^{-2}
 -\left\{4m_\pi^2 +2q^2-g_A^2\left(4m_\pi^2+3q^2\right)\right\}
 \tilde w^2 A(q)
 \right.
 \nonumber\\
 &\left.
 +g_A^2\left\{m_\pi w^2 +\tilde w^4 A(q)\right\}\right],
 \label{WCN2LO}\\
 V_S^{(3)}(q,Q)
 &=
 \frac{g_A^2}{2m_N}
 \left[
 3q^2 \tilde w^2 A(q)
 +q^2 \left\{m_\pi +w^2 A(q)\right\}\right],
 \label{VSN2LO}\\
 W_S^{(3)}(q,Q)
 &=
 -8
 \left[
 q^2 A(q)
 \left\{\left(c_4+\frac{1}{4m_N}\right)w^2
 -\frac{g_A^2}{8m_N}\left(10m_\pi^2+3q^2\right)\right\}
 \right.
 \nonumber\\
 &\left.
 +\frac{g_A^2}{8m_N}q^2 \left\{m_\pi +w^2 A(q)\right\}
 \right],
 \label{WSN2LO}
\end{align}
with the LECs $c_i$, and
\begin{align}
 A(q)
 &=
 \frac{1}{2q}\arctan\frac{q}{2m_\pi},
 \label{2piAq}\\
 \tilde w
 &=
 \sqrt{2m_\pi^2 +q^2}.
 \label{2piwtil}
\end{align}

The contact terms at N$^3$LO is characterized by
\begin{align}
 &\mathcal{M}_{\mathrm{ct;\,}\ell}^{(4)}\!\left(p,p'\right)
 \nonumber\\
 &\quad=
 \frac{1}{\pi}\left[
 \left(24D_1+\frac{3}{2}D_2-24D_5-\frac{3}{2}D_6-8D_{11}-\frac{1}{2}D_{14}\right)
 \left\{(p^2+p'^2)^2+\frac{4}{3}p^2p'^2\right\}\delta_{\ell0}\right.
 \nonumber\\
 &\quad+\left(6D_3+4D_4-6D_7-4D_8-2D_{12}-2D_{13}\right)
 \left\{(p^2+p'^2)^2-\frac{4}{3}p^2p'^2\right\}\delta_{\ell0}
 \nonumber\\
 &\quad+4\left(40D_1-\frac{5}{2}D_2+24D_5-\frac{3}{2}D_6+8D_{11}+\frac{1}{2}D_{14}\right)
 pp'(p^2+p'^2)\delta_{\ell1}
 \nonumber\\
 &\quad+\frac{8}{3}\left(24D_1+\frac{3}{2}D_2-6D_3-4D_4-24D_5-\frac{3}{2}D_6+6D_7+4D_8
 \right.
 \nonumber\\
 &\quad\left.\left.
 -8D_{11}+2D_{12}+2D_{13}-\frac{1}{2}D_{14}\right)
 p^2p'^2\delta_{\ell2}
 \right]
 \nonumber\\
 &\quad -\frac{1}{15\pi}
 \left[\left\{16(-)^{\ell}-4\right\}D_4-12D_8\right]
 \sum_{\Lambda_q=0}^2\sum_{\Lambda_Q=0}^2
 (-)^{\Lambda_Q}
 \widehat{2\!-\!\Lambda_q}\widehat{2\!-\!\Lambda_Q}
 \left[\binom{5}{2\Lambda_q}\binom{5}{2\Lambda_Q}\right]^{\!\!\frac{1}{2}}
 \nonumber\\
 &\quad\times
 \left( \Lambda_q 0 \Lambda_Q 0 | \ell 0 \right)
 \left( 2\!-\!\Lambda_q, 0, 2\!-\!\Lambda_Q, 0 | \ell 0 \right)
 \begin{Bmatrix}
  \Lambda_q   & 2\!-\!\Lambda_q & 2 \\
  2\!-\!\Lambda_Q & \Lambda_Q   & \ell
 \end{Bmatrix}
 p^{\Lambda_q+\Lambda_Q} p'^{4-\Lambda_q-\Lambda_Q}
 \nonumber\\
 &\quad +\frac{288}{\pi} D_{15}
 \sum_{L_q=0,2}
 \left(-\frac{1}{2}\right)^{\!\!L_q}
 \left( 1 0 1 0 | L_q 0 \right)^2
 \begin{Bmatrix}
  1   & 1   & 1 \\
  1   & 1   & L_q
 \end{Bmatrix}
 \nonumber\\
 &\quad\times
 \sum_{\lambda_q=0}^{\Lambda_q}\sum_{\lambda_Q=0}^{\Lambda_Q}
 (-)^{\lambda_Q}
 \widehat{L_q\!-\!\lambda_q}\widehat{L_q\!-\!\lambda_Q}
 \left[\binom{2L_q+1}{2\lambda_q}\binom{2L_q+1}{2\lambda_Q}\right]^{\!\!\frac{1}{2}}
 \nonumber\\
 &\quad\times
 \sum_{\lambda K}
 \left( \lambda_q 0 \lambda_Q 0 | \lambda 0 \right)
 \left( L_q\!-\!\lambda_q, 0, L_q\!-\!\lambda_Q, 0 | \lambda 0 \right)
 \left( \lambda 0 K 0 | \ell 0\right)^2
 \begin{Bmatrix}
  \lambda_q & L_q\!-\!\lambda_q & L_q \\
  L_q\!-\!\lambda_Q & \lambda_Q & \lambda
 \end{Bmatrix}
 \nonumber\\
 &\quad\times
 p^{\lambda_q+\lambda_Q} p'^{\,2L_q-\lambda_q-\lambda_Q}
 \left[
 \frac{1}{4}\left[\left\{\left(p^2+p'^2\right)^2 -\frac{4}{3}p^2p'^2\right\}\delta_{K0}
 -\frac{8}{3}p^2p'^2\delta_{K2}
 \right]\delta_{L_q 0}
 +\delta_{K0}\delta_{L_q 2}
 \right],
 \label{MPEN3LOct}
\end{align}
where $D_i$ are the LECs and the binomial coefficient is defined by
\begin{align}
 \binom{n_1}{n_2}
 =
 \frac{n_1!}{\left(n_1-n_2\right)!n_2!}.
 \label{binomialcoeff}
\end{align}

As regards the $2\pi$-exchange terms at N$^3$LO, 
it is convenient to classify the MPE function into
the $c_i^2$, $c_i/m_N$, $m_N^{-2}$, and two-loop (2L) terms~\cite{MACHLEIDT20111};
\begin{align}
 \mathcal{M}_{2\pi;\,\ell}^{(4)}\!\left(p,p'\right)
 =
 \mathcal{M}_{2\pi;\,\ell}^{(c_i^2)}\!\left(p,p'\right)+\mathcal{M}_{2\pi;\,\ell}^{(c_i/m_N)}\!\left(p,p'\right)
 +\mathcal{M}_{2\pi;\,\ell}^{(m_N^{-2})}\!\left(p,p'\right)+\mathcal{M}_{2\pi;\,\ell}^{(\mathrm{2L})}\!\left(p,p'\right).
 \label{MPEN3LO2pi}
\end{align}
Then, the MPE of the $c_i^2$ terms reads
\begin{align}
 \mathcal{M}_{2\pi;\,\ell}^{(c_i^2)}\!\left(p,p'\right)
 &=
 \frac{1}{2\pi^3 f_\pi^4}
 \frac{\hat{\ell}^2}{2}\int_{-1}^{1} dx P_{\ell}(x)
 \left[\left\{4(-)^{\ell}-1\right\}V_C^{(c_i^2)}(q,Q)+W_S^{(c_i^2)}(q,Q)\right],
 \label{MPEci2}\\
 V_C^{(c_i^2)}(q,Q)
 &=
 3 L(q)
 \left[\left(\frac{c_2}{6}w^2+c_3\tilde w^2 -4c_1m_\pi^2\right)^2
 +\frac{c_2^2}{45}w^4\right],
 \label{VCci2}\\
 W_S^{(c_i^2)}(q,Q)
 &=
 c_4^2 q^2 w^2 L(q),
 \label{WSci2}
\end{align}
while that of the $c_i/m_N$ terms is given by
\begin{align}
 \mathcal{M}_{2\pi;\,\ell}^{(c_i/m_N)}\!\left(p,p'\right)
 &=
 -\frac{1}{(2\pi)^3 m_N f_\pi^4}
 \frac{\hat{\ell}^2}{2}\int_{-1}^{1} dx P_{\ell}(x)
 \nonumber\\
 &\times
 \left[\left\{4(-)^{\ell}-1\right\}V_C^{(c_i/m_N)}(q,Q)+W_C^{(c_i/m_N)}(q,Q)+W_S^{(c_i/m_N)}(q,Q)\right],
 \label{MPEcimN}\\
 V_C^{(c_i/m_N)}(q,Q)
 &=
 2g_A^2
 L(q)
 \left[(c_2-6c_3)q^4
 +4(6c_1+c_2-3c_3)q^2m_\pi^2
 \right.
 \nonumber\\
 &\left.
 +6(c_2-2c_3)m_\pi^4
 +24(2c_1+c_3)m_\pi^6w^{-2}\right],
 \label{VCcimN}\\
 W_C^{(c_i/m_N)}(q,Q)
 &=
 - c_4 q^2 L(q)
 \left[g_A^2\left(8m_\pi^2+5q^2\right)+w^2\right],
 \label{WCcimN}\\
 W_S^{(c_i/m_N)}(q,Q)
 &=
 2 c_4 q^2 L(q)
 \left[g_A^2\left(16m_\pi^2+7q^2\right)-w^2\right].
 \label{WScimN}
\end{align}
The $m_N^{-2}$ terms have the MPE function of the form,
\begin{align}
 \mathcal{M}_{2\pi;\,\ell}^{(m_N^{-2})}\!\left(p,p'\right)
 &=
 -\frac{1}{(2\pi)^3 m_N^2 f_\pi^4}
 \frac{\hat{\ell}^2}{2}\int_{-1}^{1} dx P_{\ell}(x)
 \nonumber\\
 &\times
 \left[\left\{4(-)^{\ell}-1\right\}V_C^{(m_N^{-2})}(q,Q)+W_C^{(m_N^{-2})}(q,Q)+V_S^{(m_N^{-2})}(q,Q)+W_S^{(m_N^{-2})}(q,Q)\right]
 \nonumber\\
 &+\mathcal{M}_{\sigma L;\,\ell}^{(m_N^{-2})}\!\left(p,p'\right),
 \label{MPEmN-2}\\
 V_C^{(m_N^{-2})}(q,Q)
 &=
 2g_A^4
 L(q)
 \left[
 \left(2m_\pi^8 w^{-4} +8m_\pi^6w^{-2}-q^4-2m_\pi^4\right)
 +\frac{1}{2}m_\pi^6w^{-2}\right],
 \label{VCmN-2}\\
 W_C^{(m_N^{-2})}(q,Q)
 &=
 -\frac{1}{4}
  L(q)\left[
 8g_A^2\left\{\frac{3}{2}q^4+3m_\pi^2q^2+3m_\pi^4-6m_\pi^6w^{-2}
 -Q^2\left(8m_\pi^2+5q^2\right)\right\}
 \right.
 \nonumber\\
 &+4g_A^4\left\{
 Q^2\left(20m_\pi^2+7q^2-16m_\pi^4w^{-2}\right)
 \right.
 \nonumber\\
 & \left.
 +16m_\pi^8w^{-4}+12m_\pi^6w^{-2}
 -4m_\pi^4q^2w^{-2}-5q^4-6m_\pi^2q^2-6m_\pi^4
 \right\}
 -4Q^2w^2
 \bigg]
 \nonumber\\
 & -4g_A^4m_\pi^6w^{-2},
 \label{WCmN-2}\\
 V_S^{(m_N^{-2})}(q,Q)
 &=
 -4g_A^4
 q^2 L(q)
 \left[Q^2+\frac{5}{8}q^2+m_\pi^4w^{-2}\right],
 \label{VSmN-2}\\
 W_S^{(m_N^{-2})}(q,Q)
 &=
 -\frac{1}{4}
  q^2 L(q)
 \left[
 4g_A^4\left(7m_\pi^2+\frac{17}{4}q^2+4m_\pi^4w^{-2}\right)
 -32g_A^2\left(m_\pi^2+\frac{7}{16}q^2\right)+w^2
 \right],
 \label{WSmN-2}
\end{align}
\begin{align}
 \mathcal{M}_{\sigma L;\,\ell}^{(m_N^{-2})}\!\left(p,p'\right)
 &=
 \frac{9g_A^4}{\pi^3 m_N^2f_\pi^4}
 \sum_{L_q=0,2}
 \left(-\frac{1}{2}\right)^{\!\!L_q}
 \left( 1 0 1 0 | L_q 0 \right)^2
 \begin{Bmatrix}
  1   & 1   & 1 \\
  1   & 1   & L_q
 \end{Bmatrix}
 \nonumber\\
 &\times
 \sum_{\lambda_q=0}^{\Lambda_q}\sum_{\lambda_Q=0}^{\Lambda_Q}
 (-)^{\lambda_Q}
 \widehat{L_q\!-\!\lambda_q}\widehat{L_q\!-\!\lambda_Q}
 \left[\binom{2L_q+1}{2\lambda_q}\binom{2L_q+1}{2\lambda_Q}\right]^{\!\!\frac{1}{2}}
 \nonumber\\
 &\times
 \sum_{\lambda K}
 \left( \lambda_q 0 \lambda_Q 0 | \lambda 0 \right)
 \left( L_q\!-\!\lambda_q, 0, L_q\!-\!\lambda_Q, 0 | \lambda 0 \right)
 \left( \lambda 0 K 0 | \ell 0\right)^2
 \nonumber\\
 &\times
 \begin{Bmatrix}
  \lambda_q & L_q\!-\!\lambda_q & L_q \\
  L_q\!-\!\lambda_Q & \lambda_Q & \lambda
 \end{Bmatrix}
 p^{\lambda_q+\lambda_Q} p'^{\,2L_q-\lambda_q-\lambda_Q}
 \nonumber\\
 &\times
 \frac{\hat K^2}{2}
 \int_{-1}^1 dx P_K(x)
 q^{2-L_q} Q^{2-L_q} L(q).
 \label{VsigLmN-2}
\end{align}
Lastly, we write the MPE function of the 2L terms as
\begin{align}
 \mathcal{M}_{2\pi;\,\ell}^{(\mathrm{2L})}\!\left(p,p'\right)
 &=
 \frac{1}{16(2\pi)^3 f_\pi^6}
 \frac{\hat{\ell}^2}{2}\int_{-1}^{1} dx P_{\ell}(x)
 \nonumber\\
 &\times
 \left[\left\{4(-)^{\ell}-1\right\}V_C^{(\mathrm{2L})}(q,Q)+W_C^{(\mathrm{2L})}(q,Q)
 +V_S^{(\mathrm{2L})}(q,Q)+W_S^{(\mathrm{2L})}(q,Q)\right],
 \label{MPE2L}\\
 V_C^{(\mathrm{2L})}(q,Q)
 &=
 3g_A^4
 \tilde w^2 A(q)
 \left[\left(m_\pi^2+2q^2\right)\left(2m_\pi+\tilde w^2 A(q)\right)
 +4g_A^2 m_\pi \tilde w^2\right],
 \label{VC2L}\\
 W_C^{(\mathrm{2L})}(q,Q)
 &=
 -\frac{1}{6\pi^2}
 L(q) \left[
 192\pi^2f_\pi^2w^2\bar d_3
 \left\{2g_A^2\tilde w^2-\frac{3}{5}\left(g_A^2-1\right) w^2\right\}
 \right.
 \nonumber\\
 & +\left\{6g_A^2\tilde w^2-\left(g_A^2-1\right)w^2\right\}
 \nonumber\\
 &\times \Big[
 384\pi^2f_\pi^2
 \left\{\tilde w^2 \left(\bar d_1 +\bar d_2\right)+4m_\pi^2 \bar d_5 \right\}
 +L(q)\left\{4m_\pi^2\left(1+2g_A^2\right)+q^2\left(1+5g_A^2\right)\right\}
 \nonumber\\
 & \left.
 -\frac{1}{3}q^2\left(5+13g_A^2\right)
 -8m_\pi^2\left(1+2g_A^2\right)
 \Big]\right],
 \label{WC2L}\\
 V_S^{(\mathrm{2L})}(q,Q)
 &=
 -64g_A^2 f_\pi^2
 \left(\bar d_{14}-\bar d_{15}\right)
 q^2 w^2 L(q),
 \label{VS2L}\\
 W_S^{(\mathrm{2L})}(q,Q)
 &=
 3g_A^4
 q^2 w^2 A(q)
 \left[ w^2 A(q) +2m_\pi\left(1+2g_A^2\right) \right],
 \label{WS2L}
\end{align}
where $\bar d_{i}$ are the LECs.

\section{Constants}
\label{SecConstants}
The constants necessary for the ME calculations are
the average nucleon mass $m_N$,
the average pion mass $m_\pi$,
the pion-decay constant $f_\pi$,
the axial vector coupling constant $g_A$,
and LECs $\left(C_S, C_T, C_i, D_i, c_i,~\mathrm{and}~\bar d_i \right)$,
as well as the regulator parameters (the exponent $n$ and the cutoff $\Lambda$).
These constants are listed in Tables~\ref{tableconst1} and \ref{tableconst2}.
We employ natural units in which one assumes $\hbar=1$ and $c=1$.

The LECs of $\Lambda=500$ and 600 MeV are taken from Ref.~\cite{MACHLEIDT20111}.
Note that the LECs $D_i$ listed in Table F.1 of Ref.~\cite{MACHLEIDT20111} are slightly modified because $n=2$ is
now used consistently for all fourth order contact terms~\cite{MachleidtPrivate}.
The new values of $D_i$ are given in Table~\ref{tableconst2}.
We also test the lower-cutoff parameters, $\Lambda=450$~MeV~\cite{PhysRevC.87.014322,PhysRevC.89.044321},
which are now represented by the partial-wave-independent expression~\cite{MachleidtPrivate}.
\begin{table}[!h]
 \caption{The average masses ($m_N$ and $m_\pi$) and the pion-decay constant $f_\pi$ are in units of MeV, 
 while the axial vector coupling constant $g_A$ is dimensionless.}
 \label{tableconst1}
 \begin{center}
   \begin{tabular*}{0.45\textwidth}{@{\extracolsep{\fill}}cccc}
    \hline
    $m_N$      & $m_\pi$    & $f_\pi$ & $g_A$\\
    $938.9187$ & $138.0394$ & $92.4$  & $1.29 $\\
    \hline
   \end{tabular*}
 \end{center}
\end{table}
\begin{longtable}{clD{.}{.}{6}D{.}{.}{6}D{.}{.}{6}}
 \caption{The regulator parameter $n$ and the LECs for three sets of $\Lambda$.
 The LECs $c_i$ and $\bar d_i$ involved in the $2\pi$-exchange terms
 are respectively in units of GeV$^{-1}$ and GeV$^{-2}$,
 while $C_S$ and $C_T$ at LO having the charge dependence
 are in units of $10^4$~GeV$^{-2}$. The LECs $C_i$ at NLO and $D_i$ at
 N$^3$LO are in units of $10^4$~GeV$^{-4}$ and $10^4$~GeV$^{-6}$, respectively.}
 \label{tableconst2}\\
 \endfirsthead
 \caption*{Table~\ref{tableconst2} continued.}\\
 \hline
   & & \multicolumn{1}{c}{$\Lambda=450$ MeV} & 
       \multicolumn{1}{c}{$\Lambda=500$ MeV} & 
       \multicolumn{1}{c}{$\Lambda=600$ MeV} \\ \cline{3-5}
 \endhead
   \hline
   & & \multicolumn{1}{c}{$\Lambda=450$ MeV} & 
       \multicolumn{1}{c}{$\Lambda=500$ MeV} & 
       \multicolumn{1}{c}{$\Lambda=600$ MeV} \\ \cline{3-5}
   \multirow{3}{*}{$n$} & $1\pi$ term     & 4 & 4 & 4 \\
			& LO contacts	  & 3 & 3 & 3 \\
			& All other terms & 3 & 2 & 2 \\
    $c_1$		      & & -0.81 & -0.81 & -0.81 \\
    $c_2$		      & &  3.28 &  2.80 &  2.80 \\
    $c_3$		      & & -3.40 & -3.20 & -3.20 \\
    $c_4$		      & &  3.40 &  5.40 &  5.40 \\
    $\bar d_1+\bar d_2$	      & &  3.06 &  3.06 &  3.06 \\
    $\bar d_3$		      & & -3.27 & -3.27 & -3.27 \\
    $\bar d_5$		      & &  0.45 &  0.45 &  0.45 \\
    $\bar d_{14}-\bar d_{15}$ & & -5.65 & -5.65 & -5.65 \\
    $C_S^{pp}$ & & -0.011603 & -0.009991 & -0.009943 \\
    $C_S^{nn}$ & & -0.011610 & -0.010011 & -0.009949 \\
    $C_S^{np}$ & & -0.011619 & -0.010028 & -0.009955 \\
    $C_T^{pp}$ & &  0.000229 &  0.000523 &  0.000695 \\
    $C_T^{nn}$ & &  0.000236 &	0.000543 &  0.000701 \\
    $C_T^{np}$ & &  0.000245 &	0.000561 &  0.000707 \\
    $C_1$      & &  0.055936 &	0.051949 &  0.046433 \\
    $C_2$      & &  0.092976 &	0.163034 &  0.174354 \\
    $C_3$      & & -0.003346 &	0.003249 &  0.006562 \\
    $C_4$      & & -0.045017 & -0.048954 & -0.050066 \\
    $C_5$      & & -0.072187 & -0.075081 & -0.086978 \\
    $C_6$      & & -0.004651 & -0.013343 & -0.013639 \\
    $C_7$      & & -0.173839 & -0.225500 & -0.214188 \\
    $D_1$      & &  0.027479 & -0.016761 & -0.026710 \\
    $D_2$      & &  2.573138 &	2.478838 &  3.054553 \\
    $D_3$      & &  1.773384 &	0.915937 &  0.100079 \\
    $D_4$      & & -1.735485 & -0.810894 & -0.047571 \\
    $D_5$      & &  0.132598 &	0.144506 &  0.131379 \\
    $D_6$      & &  1.810399 &	1.324597 &  1.324064 \\
    $D_7$      & &  0.608374 &	0.103610 &  0.236467 \\
    $D_8$      & & -0.512382 & -0.189115 & -0.268515 \\
    $D_9$      & & -0.462892 & -0.547244 & -0.633636 \\
    $D_{10}$   & &  1.717083 &	2.502115 &  2.152571 \\
    $D_{11}$   & & -0.123624 & -0.129033 & -0.154661 \\
    $D_{12}$   & & -0.069623 &	0.110915 & -0.121515 \\
    $D_{13}$   & & -0.101480 & -0.012318 & -0.010343 \\
    $D_{14}$   & & -0.916369 & -1.368423 & -0.860344 \\
    $D_{15}$   & & -0.051120 &	0.147177 & -0.013131 \\
   \hline
\end{longtable}

\clearpage
\bibliographystyle{iopart-num}
\bibliography{paperRCM}

\providecommand{\newblock}{}
\begin{thebibliography}{10}
\expandafter\ifx\csname url\endcsname\relax
  \def\url#1{{\tt #1}}\fi
\expandafter\ifx\csname urlprefix\endcsname\relax\def\urlprefix{URL }\fi
\providecommand{\eprint}[2][]{\url{#2}}

\bibitem{10.1143/PTP.44.646}
Band\={o} H, Nagata S and Yamamoto Y 1970 {\em Prog. Theor. Phys.\/} {\bf 44}
  646--662 ISSN 0033-068X \urlprefix\url{https://doi.org/10.1143/PTP.44.646}

\bibitem{10.1143/PTP.45.1515}
Band\={o} H, Nagata S and Yamamoto Y 1971 {\em Prog. Theor. Phys.\/} {\bf 45}
  1515--1526 ISSN 0033-068X \urlprefix\url{https://doi.org/10.1143/PTP.45.1515}

\bibitem{10.1143/PTP.59.817}
Yamamoto Y, Band\={o} H and Nagata S 1978 {\em Prog. Theor. Phys.\/} {\bf 59}
  817--830 ISSN 0033-068X \urlprefix\url{https://doi.org/10.1143/PTP.59.817}

\bibitem{10.1143/PTP.117.189}
Togashi T and Kat\={o} K 2007 {\em Prog. Theor. Phys.\/} {\bf 117} 189--194
  ISSN 0033-068X \urlprefix\url{https://doi.org/10.1143/PTP.117.189}

\bibitem{10.1143/PTP.121.299}
Togashi T, Murakami T and Kat\={o} K 2009 {\em Prog. Theor. Phys.\/} {\bf 121}
  299--317 ISSN 0033-068X \urlprefix\url{https://doi.org/10.1143/PTP.121.299}

\bibitem{10.1143/PTP.124.315}
Yamamoto Y, Togashi T and Kat\={o} K 2010 {\em Prog. Theor. Phys.\/} {\bf 124}
  315--330 ISSN 0033-068X \urlprefix\url{https://doi.org/10.1143/PTP.124.315}

\bibitem{Kato12}
Togashi T and Kat\={o} K 2012 {\em {Antisymmetrized Molecular Dynamics with
  Bare Nuclear Interactions: Brueckner-AMD, and Its Applications to Light
  Nuclei}\/} (Rijeka: IntechOpen) \urlprefix\url{https://doi.org/10.5772/36510}

\bibitem{PhysRevC.62.014001}
Wiringa R~B, Pieper S~C, Carlson J and Pandharipande V~R 2000 {\em Phys. Rev.
  C\/} {\bf 62}(1) 014001
  \urlprefix\url{https://link.aps.org/doi/10.1103/PhysRevC.62.014001}

\bibitem{RevModPhys.87.1067}
Carlson J, Gandolfi S, Pederiva F, Pieper S~C, Schiavilla R, Schmidt K~E and
  Wiringa R~B 2015 {\em Rev. Mod. Phys.\/} {\bf 87}(3) 1067--1118
  \urlprefix\url{https://link.aps.org/doi/10.1103/RevModPhys.87.1067}

\bibitem{PhysRevLett.111.062502}
Datar V~M, Chakrabarty D~R, Kumar S, Nanal V, Pastore S, Wiringa R~B, Behera
  S~P, Chatterjee A, Jenkins D, Lister C~J, Mirgule E~T, Mitra A, Pillay R~G,
  Ramachandran K, Roberts O~J, Rout P~C, Shrivastava A and Sugathan P 2013 {\em
  Phys. Rev. Lett.\/} {\bf 111}(6) 062502
  \urlprefix\url{https://link.aps.org/doi/10.1103/PhysRevLett.111.062502}

\bibitem{PhysRevLett.84.5728}
Navr\'atil P, Vary J~P and Barrett B~R 2000 {\em Phys. Rev. Lett.\/} {\bf
  84}(25) 5728--5731
  \urlprefix\url{https://link.aps.org/doi/10.1103/PhysRevLett.84.5728}

\bibitem{PhysRevC.62.054311}
Navr\'atil P, Vary J~P and Barrett B~R 2000 {\em Phys. Rev. C\/} {\bf 62}(5)
  054311 \urlprefix\url{https://link.aps.org/doi/10.1103/PhysRevC.62.054311}

\bibitem{PhysRevC.86.054301}
Abe T, Maris P, Otsuka T, Shimizu N, Utsuno Y and Vary J~P 2012 {\em Phys. Rev.
  C\/} {\bf 86}(5) 054301
  \urlprefix\url{https://link.aps.org/doi/10.1103/PhysRevC.86.054301}

\bibitem{10.1093/ptep/pts012}
Shimizu N, Abe T, Tsunoda Y, Utsuno Y, Yoshida T, Mizusaki T, Honma M and
  Otsuka T 2012 {\em Prog. Theor. Exp. Phys.\/} {\bf 2012} 01A205 ISSN
  2050-3911 \urlprefix\url{https://doi.org/10.1093/ptep/pts012}

\bibitem{PhysRevLett.114.212502}
Hupin G, Quaglioni S and Navr\'atil P 2015 {\em Phys. Rev. Lett.\/} {\bf
  114}(21) 212502
  \urlprefix\url{https://link.aps.org/doi/10.1103/PhysRevLett.114.212502}

\bibitem{PhysRevLett.119.062501}
Kravvaris K and Volya A 2017 {\em Phys. Rev. Lett.\/} {\bf 119}(6) 062501
  \urlprefix\url{https://link.aps.org/doi/10.1103/PhysRevLett.119.062501}

\bibitem{PhysRevLett.98.162503}
Dytrych T, Sviratcheva K~D, Bahri C, Draayer J~P and Vary J~P 2007 {\em Phys.
  Rev. Lett.\/} {\bf 98}(16) 162503
  \urlprefix\url{https://link.aps.org/doi/10.1103/PhysRevLett.98.162503}

\bibitem{PhysRevLett.111.252501}
Dytrych T, Launey K~D, Draayer J~P, Maris P, Vary J~P, Saule E, Catalyurek U,
  Sosonkina M, Langr D and Caprio M~A 2013 {\em Phys. Rev. Lett.\/} {\bf
  111}(25) 252501
  \urlprefix\url{https://link.aps.org/doi/10.1103/PhysRevLett.111.252501}

\bibitem{PhysRevC.102.044608}
Dreyfuss A~C, Launey K~D, Escher J~E, Sargsyan G~H, Baker R~B, Dytrych T and
  Draayer J~P 2020 {\em Phys. Rev. C\/} {\bf 102}(4) 044608
  \urlprefix\url{https://link.aps.org/doi/10.1103/PhysRevC.102.044608}

\bibitem{PhysRevLett.113.032503}
Romero-Redondo C, Quaglioni S, Navr\'atil P and Hupin G 2014 {\em Phys. Rev.
  Lett.\/} {\bf 113}(3) 032503
  \urlprefix\url{https://link.aps.org/doi/10.1103/PhysRevLett.113.032503}

\bibitem{kravvaris2020ab}
Kravvaris K, Quaglioni S, Hupin G and Navratil P 2020 Ab initio framework for
  nuclear scattering and reactions induced by light projectiles
  (\textit{Preprint} \eprint{2012.00228})

\bibitem{PhysRevLett.117.222501}
Romero-Redondo C, Quaglioni S, Navr\'atil P and Hupin G 2016 {\em Phys. Rev.
  Lett.\/} {\bf 117}(22) 222501
  \urlprefix\url{https://link.aps.org/doi/10.1103/PhysRevLett.117.222501}

\bibitem{NEFF2004357}
Neff T and Feldmeier H 2004 {\em Nucl. Phys. A\/} {\bf 738} 357 -- 361 ISSN
  0375-9474
  \urlprefix\url{http://www.sciencedirect.com/science/article/pii/S0375947404006025}

\bibitem{PhysRevLett.98.032501}
Chernykh M, Feldmeier H, Neff T, von Neumann-Cosel P and Richter A 2007 {\em
  Phys. Rev. Lett.\/} {\bf 98}(3) 032501
  \urlprefix\url{https://link.aps.org/doi/10.1103/PhysRevLett.98.032501}

\bibitem{Borasoy2007-1}
Borasoy B, Epelbaum E, Krebs H, Lee D and Mei{\ss}ner U~G 2007 {\em Eur. Phys.
  J. A\/} {\bf 31} 105--123 ISSN 1434-601X
  \urlprefix\url{https://doi.org/10.1140/epja/i2006-10154-1}

\bibitem{Borasoy2007-2}
Borasoy B, Epelbaum E, Krebs H, Lee D and Mei{\ss}ner U~G 2007 {\em Eur. Phys.
  J. A\/} {\bf 34} 185--196 ISSN 1434-601X
  \urlprefix\url{https://doi.org/10.1140/epja/i2007-10500-9}

\bibitem{PhysRevLett.106.192501}
Epelbaum E, Krebs H, Lee D and Mei\ss{}ner U~G 2011 {\em Phys. Rev. Lett.\/}
  {\bf 106}(19) 192501
  \urlprefix\url{https://link.aps.org/doi/10.1103/PhysRevLett.106.192501}

\bibitem{PhysRevLett.109.252501}
Epelbaum E, Krebs H, L\"ahde T~A, Lee D and Mei\ss{}ner U~G 2012 {\em Phys.
  Rev. Lett.\/} {\bf 109}(25) 252501
  \urlprefix\url{https://link.aps.org/doi/10.1103/PhysRevLett.109.252501}

\bibitem{Epelbaum2013}
Epelbaum E, Krebs H, L{\"a}hde T~A, Lee D and Mei{\ss}ner U~G 2013 {\em Eur.
  Phys. J. A\/} {\bf 49} 82 ISSN 1434-601X
  \urlprefix\url{https://doi.org/10.1140/epja/i2013-13082-y}

\bibitem{PhysRevLett.112.102501}
Epelbaum E, Krebs H, L\"ahde T~A, Lee D, Mei\ss{}ner U~G and Rupak G 2014 {\em
  Phys. Rev. Lett.\/} {\bf 112}(10) 102501
  \urlprefix\url{https://link.aps.org/doi/10.1103/PhysRevLett.112.102501}

\bibitem{Elhatisari2015}
Elhatisari S, Lee D, Rupak G, Epelbaum E, Krebs H, L{\"a}hde T~A, Luu T and
  Mei{\ss}ner U~G 2015 {\em Nature\/} {\bf 528} 111--114
  \urlprefix\url{https://doi.org/10.1038/nature16067}

\bibitem{PhysRevLett.117.132501}
Elhatisari S, Li N, Rokash A, Alarc\'on J~M, Du D, Klein N, Lu B~n, Mei\ss{}ner
  U~G, Epelbaum E, Krebs H, L\"ahde T~A, Lee D and Rupak G 2016 {\em Phys. Rev.
  Lett.\/} {\bf 117}(13) 132501
  \urlprefix\url{https://link.aps.org/doi/10.1103/PhysRevLett.117.132501}

\bibitem{PhysRevLett.119.222505}
Elhatisari S, Epelbaum E, Krebs H, L\"ahde T~A, Lee D, Li N, Lu B~n,
  Mei\ss{}ner U~G and Rupak G 2017 {\em Phys. Rev. Lett.\/} {\bf 119}(22)
  222505
  \urlprefix\url{https://link.aps.org/doi/10.1103/PhysRevLett.119.222505}

\bibitem{RevModPhys.90.035004}
Freer M, Horiuchi H, Kanada-En'yo Y, Lee D and Mei\ss{}ner U~G 2018 {\em Rev.
  Mod. Phys.\/} {\bf 90}(3) 035004
  \urlprefix\url{https://link.aps.org/doi/10.1103/RevModPhys.90.035004}

\bibitem{10.1143/PTP.27.793}
Shimodaya I, Tamagaki R and Tanaka H 1962 {\em Prog. Theor. Phys.\/} {\bf 27}
  793--810 ISSN 0033-068X \urlprefix\url{https://doi.org/10.1143/PTP.27.793}

\bibitem{Weinberg1979327}
Weinberg S 1979 {\em Phys. A\/} {\bf 96} 327 -- 340 ISSN 0378-4371
  \urlprefix\url{http://www.sciencedirect.com/science/article/pii/0378437179902231}

\bibitem{Epelbaum2006654}
Epelbaum E 2006 {\em Prog. Part. Nucl. Phys.\/} {\bf 57} 654 -- 741 ISSN
  0146-6410
  \urlprefix\url{http://www.sciencedirect.com/science/article/pii/S0146641005001018}

\bibitem{MACHLEIDT20111}
Machleidt R and Entem D~R 2011 {\em Phys. Rep.\/} {\bf 503} 1 -- 75 ISSN
  0370-1573
  \urlprefix\url{http://www.sciencedirect.com/science/article/pii/S0370157311000457}

\bibitem{10.1143/PTPS.52.1}
Ikeda K, Marumori T, Tamagaki R and Tanaka H 1972 {\em Prog. Theor. Phys.
  Suppl.\/} {\bf 52} 1--24 ISSN 0375-9687
  \urlprefix\url{https://doi.org/10.1143/PTPS.52.1}

\bibitem{10.1143/PTPS.52.25}
Hiura J and Tamagaki R 1972 {\em Prog. Theor. Phys. Suppl.\/} {\bf 52} 25--88
  ISSN 0375-9687 \urlprefix\url{https://doi.org/10.1143/PTPS.52.25}

\bibitem{VOLKOV196533}
Volkov A 1965 {\em Nucl. Phys.\/} {\bf 74} 33 -- 58 ISSN 0029-5582
  \urlprefix\url{http://www.sciencedirect.com/science/article/pii/0029558265902440}

\bibitem{brink1966proc}
Brink D~M 1966 {\em {The Alpha-Particle Model of Light Nuclei}\/} (New York and
  London: Academic Press)

\bibitem{ENTEM200293}
Entem D~R and Machleidt R 2002 {\em Phys. Lett. B\/} {\bf 524} 93 -- 98 ISSN
  0370-2693
  \urlprefix\url{http://www.sciencedirect.com/science/article/pii/S0370269301013636}

\bibitem{PhysRevC.66.014002}
Entem D~R and Machleidt R 2002 {\em Phys. Rev. C\/} {\bf 66}(1) 014002
  \urlprefix\url{https://link.aps.org/doi/10.1103/PhysRevC.66.014002}

\bibitem{PhysRevC.68.041001}
Entem D~R and Machleidt R 2003 {\em Phys. Rev. C\/} {\bf 68}(4) 041001
  \urlprefix\url{https://link.aps.org/doi/10.1103/PhysRevC.68.041001}

\bibitem{PhysRev.89.1102}
Hill D~L and Wheeler J~A 1953 {\em Phys. Rev.\/} {\bf 89}(5) 1102--1145
  \urlprefix\url{https://link.aps.org/doi/10.1103/PhysRev.89.1102}

\bibitem{PhysRev.108.311}
Griffin J~J and Wheeler J~A 1957 {\em Phys. Rev.\/} {\bf 108}(2) 311--327
  \urlprefix\url{https://link.aps.org/doi/10.1103/PhysRev.108.311}

\bibitem{10.1143/PTP.56.583}
Mito Y and Kamimura M 1976 {\em Prog. Theor. Phys.\/} {\bf 56} 583--598 ISSN
  0033-068X \urlprefix\url{https://doi.org/10.1143/PTP.56.583}

\bibitem{10.1143/PTPS.62.236}
Kamimura M 1977 {\em Prog. Theor. Phys. Suppl.\/} {\bf 62} 236--294 ISSN
  0375-9687 \urlprefix\url{https://doi.org/10.1143/PTPS.62.236}

\bibitem{PhysRevC.98.054306}
Matsuno H, Kanada-En'yo Y and Itagaki N 2018 {\em Phys. Rev. C\/} {\bf 98}(5)
  054306 \urlprefix\url{https://link.aps.org/doi/10.1103/PhysRevC.98.054306}

\bibitem{10.1093/ptep/ptz046}
Itagaki N, Matsuno H and Kanada-En'o Y 2019 {\em Prog. Theor. Exp. Phys.\/}
  {\bf 2019} 063D02 ISSN 2050-3911
  \urlprefix\url{https://doi.org/10.1093/ptep/ptz046}

\bibitem{ANGELI201369}
Angeli I and Marinova K 2013 {\em At. Data Nucl. Data Tables\/} {\bf 99} 69 --
  95 ISSN 0092-640X
  \urlprefix\url{http://www.sciencedirect.com/science/article/pii/S0092640X12000265}

\bibitem{BORISYUK201059}
Borisyuk D 2010 {\em Nucl. Phys. A\/} {\bf 843} 59 -- 67 ISSN 0375-9474
  \urlprefix\url{http://www.sciencedirect.com/science/article/pii/S0375947410005117}

\bibitem{PhysRevC.87.014322}
Coraggio L, Holt J~W, Itaco N, Machleidt R and Sammarruca F 2013 {\em Phys.
  Rev. C\/} {\bf 87}(1) 014322
  \urlprefix\url{https://link.aps.org/doi/10.1103/PhysRevC.87.014322}

\bibitem{PhysRevC.89.044321}
Coraggio L, Holt J~W, Itaco N, Machleidt R, Marcucci L~E and Sammarruca F 2014
  {\em Phys. Rev. C\/} {\bf 89}(4) 044321
  \urlprefix\url{https://link.aps.org/doi/10.1103/PhysRevC.89.044321}

\bibitem{Wang_2021}
Wang M, Huang W, Kondev F, Audi G and Naimi S 2021 {\em Chin. Phys. C\/} {\bf
  45} 030003 \urlprefix\url{https://doi.org/10.1088/1674-1137/abddaf}

\bibitem{TILLEY2004155}
Tilley D, Kelley J, Godwin J, Millener D, Purcell J, Sheu C and Weller H 2004
  {\em Nucl. Phys. A\/} {\bf 745} 155 -- 362 ISSN 0375-9474
  \urlprefix\url{http://www.sciencedirect.com/science/article/pii/S0375947404010267}

\bibitem{itagaki2020challenge}
Itagaki N, Fukui T and Tohsaki A 2020 {Challenge for describing the cluster
  states starting with realistic interaction} (\textit{Preprint}
  \eprint{2003.08546})

\bibitem{PhysRevC.49.1814}
Tohsaki A 1994 {\em Phys. Rev. C\/} {\bf 49}(4) 1814--1817
  \urlprefix\url{https://link.aps.org/doi/10.1103/PhysRevC.49.1814}

\bibitem{PhysRev.104.123}
Heydenburg N~P and Temmer G~M 1956 {\em Phys. Rev.\/} {\bf 104}(1) 123--134
  \urlprefix\url{https://link.aps.org/doi/10.1103/PhysRev.104.123}

\bibitem{PhysRev.104.135}
Russell J~L, Phillips G~C and Reich C~W 1956 {\em Phys. Rev.\/} {\bf 104}(1)
  135--142 \urlprefix\url{https://link.aps.org/doi/10.1103/PhysRev.104.135}

\bibitem{PhysRev.109.850}
Nilson R, Jentschke W~K, Briggs G~R, Kerman R~O and Snyder J~N 1958 {\em Phys.
  Rev.\/} {\bf 109}(3) 850--860
  \urlprefix\url{https://link.aps.org/doi/10.1103/PhysRev.109.850}

\bibitem{PhysRev.117.525}
Jones C~M, Phillips G~C and Miller P~D 1960 {\em Phys. Rev.\/} {\bf 117}(2)
  525--530 \urlprefix\url{https://link.aps.org/doi/10.1103/PhysRev.117.525}

\bibitem{MiyakeBICR1961}
Miyake K 1961 {\em Bulletin of the Institute for Chemical Research, Kyoto
  University\/} {\bf 39} 313 \urlprefix\url{http://hdl.handle.net/2433/75850}

\bibitem{PhysRev.129.2252}
Tombrello T~A and Senhouse L~S 1963 {\em Phys. Rev.\/} {\bf 129}(5) 2252--2258
  \urlprefix\url{https://link.aps.org/doi/10.1103/PhysRev.129.2252}

\bibitem{PhysRevLett.29.1331}
Bacher A~D, Resmini F~G, Conzett H~E, de~Swiniarski R, Meiner H and Ernst J
  1972 {\em Phys. Rev. Lett.\/} {\bf 29}(19) 1331--1333
  \urlprefix\url{https://link.aps.org/doi/10.1103/PhysRevLett.29.1331}

\bibitem{10.1143/PTP.53.677}
Tanabe F, Tohsaki A and Tamagaki R 1975 {\em Prog. Theor. Phys.\/} {\bf 53}
  677--691 ISSN 0033-068X \urlprefix\url{https://doi.org/10.1143/PTP.53.677}

\bibitem{10.1143/PTP.45.1786}
Hasegawa A and Nagata S 1971 {\em Prog. Theor. Phys.\/} {\bf 45} 1786--1807
  ISSN 0033-068X \urlprefix\url{https://doi.org/10.1143/PTP.45.1786}

\bibitem{10.1143/PTPS.E68.242}
Tamagaki R 1968 {\em Prog. Theor. Phys. Suppl.\/} {\bf E68} 242--258 ISSN
  0375-9687 \urlprefix\url{https://doi.org/10.1143/PTPS.E68.242}

\bibitem{10.1143/PTPS.62.11}
Saito S 1977 {\em Prog. Theor. Phys. Suppl.\/} {\bf 62} 11--89 ISSN 0375-9687
  \urlprefix\url{https://doi.org/10.1143/PTPS.62.11}

\bibitem{10.1143/PTPS.62.90}
Horiuchi H 1977 {\em Prog. Theor. Phys. Suppl.\/} {\bf 62} 90--190 ISSN
  0375-9687 \urlprefix\url{https://doi.org/10.1143/PTPS.62.90}

\bibitem{PhysRevC.93.034606}
Fukui T, Taniguchi Y, Suhara T, Kanada-En'yo Y and Ogata K 2016 {\em Phys. Rev.
  C\/} {\bf 93}(3) 034606
  \urlprefix\url{https://link.aps.org/doi/10.1103/PhysRevC.93.034606}

\bibitem{Machleidt1989}
Machleidt R 1989 {\em The Meson Theory of Nuclear Forces and Nuclear
  Structure\/} (Boston, MA: Springer US) pp 189--376 ISBN 978-1-4613-9907-0
  \urlprefix\url{https://doi.org/10.1007/978-1-4613-9907-0_2}

\bibitem{PhysRevC.63.024001}
Machleidt R 2001 {\em Phys. Rev. C\/} {\bf 63}(2) 024001
  \urlprefix\url{https://link.aps.org/doi/10.1103/PhysRevC.63.024001}

\bibitem{PhysRevC.49.2932}
van Kolck U 1994 {\em Phys. Rev. C\/} {\bf 49}(6) 2932--2941
  \urlprefix\url{https://link.aps.org/doi/10.1103/PhysRevC.49.2932}

\bibitem{EpelbaumPhysRevC.66.064001}
Epelbaum E, Nogga A, Gl\"ockle W, Kamada H, Mei\ss{}ner U~G and Wita\l{}a H
  2002 {\em Phys. Rev. C\/} {\bf 66}(6) 064001
  \urlprefix\url{http://link.aps.org/doi/10.1103/PhysRevC.66.064001}

\bibitem{PhysRevC.87.014327}
Maris P, Vary J~P and Navr\'atil P 2013 {\em Phys. Rev. C\/} {\bf 87}(1) 014327
  \urlprefix\url{https://link.aps.org/doi/10.1103/PhysRevC.87.014327}

\bibitem{NavratilFS2007}
Navr{\'a}til P 2007 {\em Few-Body Syst.\/} {\bf 41} 117--140 ISSN 1432-5411
  \urlprefix\url{http://dx.doi.org/10.1007/s00601-007-0193-3}

\bibitem{PhysRevC.98.044305}
Fukui T, De~Angelis L, Ma Y~Z, Coraggio L, Gargano A, Itaco N and Xu F~R 2018
  {\em Phys. Rev. C\/} {\bf 98}(4) 044305
  \urlprefix\url{https://link.aps.org/doi/10.1103/PhysRevC.98.044305}

\bibitem{MachleidtPrivate}
Machleidt R 2020 private communication

\end{thebibliography}

\end{document}